\PassOptionsToPackage{unicode}{hyperref}
\PassOptionsToPackage{hyphens}{url}
\PassOptionsToPackage{dvipsnames,svgnames,x11names}{xcolor}
\documentclass[
  12pt]{article}

\usepackage{amsmath,amssymb}
\usepackage{iftex}
\ifPDFTeX
  \usepackage[T1]{fontenc}
  \usepackage[utf8]{inputenc}
  \usepackage{textcomp} 
\else 
  \usepackage{unicode-math}
  \defaultfontfeatures{Scale=MatchLowercase}
  \defaultfontfeatures[\rmfamily]{Ligatures=TeX,Scale=1}
\fi
\usepackage{lmodern}
\ifPDFTeX\else  
\fi
\IfFileExists{upquote.sty}{\usepackage{upquote}}{}
\IfFileExists{microtype.sty}{
  \usepackage[]{microtype}
  \UseMicrotypeSet[protrusion]{basicmath} 
}{}
\makeatletter
\@ifundefined{KOMAClassName}{
  \IfFileExists{parskip.sty}{%
    \usepackage{parskip}
  }{
    \setlength{\parindent}{0pt}
    \setlength{\parskip}{6pt plus 2pt minus 1pt}}
}{
  \KOMAoptions{parskip=half}}
\makeatother
\usepackage{xcolor}
\makeatletter
\ifx\paragraph\undefined\else
  \let\oldparagraph\paragraph
  \renewcommand{\paragraph}{
    \@ifstar
      \xxxParagraphStar
      \xxxParagraphNoStar
  }
  \newcommand{\xxxParagraphStar}[1]{\oldparagraph*{#1}\mbox{}}
  \newcommand{\xxxParagraphNoStar}[1]{\oldparagraph{#1}\mbox{}}
\fi
\ifx\subparagraph\undefined\else
  \let\oldsubparagraph\subparagraph
  \renewcommand{\subparagraph}{
    \@ifstar
      \xxxSubParagraphStar
      \xxxSubParagraphNoStar
  }
  \newcommand{\xxxSubParagraphStar}[1]{\oldsubparagraph*{#1}\mbox{}}
  \newcommand{\xxxSubParagraphNoStar}[1]{\oldsubparagraph{#1}\mbox{}}
\fi
\makeatother

\usepackage{longtable,booktabs,array}
\usepackage{calc} 
\usepackage{etoolbox}
\makeatletter
\patchcmd\longtable{\par}{\if@noskipsec\mbox{}\fi\par}{}{}
\makeatother
\IfFileExists{footnotehyper.sty}{\usepackage{footnotehyper}}{\usepackage{footnote}}
\makesavenoteenv{longtable}
\usepackage{graphicx}
\makeatletter
\def\maxwidth{\ifdim\Gin@nat@width>\linewidth\linewidth\else\Gin@nat@width\fi}
\def\maxheight{\ifdim\Gin@nat@height>\textheight\textheight\else\Gin@nat@height\fi}
\makeatother
\setkeys{Gin}{width=\maxwidth,height=\maxheight,keepaspectratio}
\makeatletter
\def\fps@figure{htbp}
\makeatother

\makeatletter
\@ifpackageloaded{caption}{}{\usepackage{caption}}
\AtBeginDocument{%
\ifdefined\contentsname
  \renewcommand*\contentsname{Table of contents}
\else
  \newcommand\contentsname{Table of contents}
\fi
\ifdefined\listfigurename
  \renewcommand*\listfigurename{List of Figures}
\else
  \newcommand\listfigurename{List of Figures}
\fi
\ifdefined\listtablename
  \renewcommand*\listtablename{List of Tables}
\else
  \newcommand\listtablename{List of Tables}
\fi
\ifdefined\figurename
  \renewcommand*\figurename{Figure}
\else
  \newcommand\figurename{Figure}
\fi
\ifdefined\tablename
  \renewcommand*\tablename{Table}
\else
  \newcommand\tablename{Table}
\fi
}
\@ifpackageloaded{float}{}{\usepackage{float}}
\floatstyle{ruled}
\@ifundefined{c@chapter}{\newfloat{codelisting}{h}{lop}}{\newfloat{codelisting}{h}{lop}[chapter]}
\floatname{codelisting}{Listing}

\makeatother
\makeatletter
\@ifpackageloaded{caption}{}{\usepackage{caption}}
\@ifpackageloaded{subcaption}{}{\usepackage{subcaption}}
\makeatother

\ifLuaTeX
  \usepackage{selnolig}  
\fi
\usepackage[]{natbib}
\usepackage{bookmark}

\IfFileExists{xurl.sty}{\usepackage{xurl}}{} 
\hypersetup{
  pdftitle={Title},
  pdfauthor={Author 1; Author 2},
  pdfkeywords={3 to 6 keywords, that do not appear in the title},
  colorlinks=true,
  linkcolor={blue},
  filecolor={Maroon},
  citecolor={Blue},
  urlcolor={Blue},
  pdfcreator={LaTeX via pandoc}}

\newcommand{\anon}{1}

\usepackage{xcolor}
\usepackage{amsmath}
\usepackage{graphicx}
\usepackage{enumerate}
\usepackage{natbib}
\usepackage{url} 
\usepackage{subcaption} 

\RequirePackage{amsmath,amsfonts,amssymb,amsthm}
\usepackage{authblk}
\usepackage{makecell} 
\usepackage{float}
\PassOptionsToPackage{unicode,colorlinks,bookmarksopen,bookmarksnumbered,citecolor=blue,urlcolor=blue,linkcolor=blue}{hyperref}
\usepackage[normalem]{ulem}
\usepackage{graphicx,bookmark}
\usepackage{booktabs,mathtools}
\usepackage{rotating}
\usepackage{here}
\usepackage{dsfont}
\usepackage{booktabs}
\usepackage{multirow}
\numberwithin{equation}{section}

\newtheorem{theorem}{Theorem}
\newtheorem{corollary}{Corollary}
\newtheorem{lemma}{Lemma}

\newtheorem{proposition}{Proposition}
\newtheorem{remark}{Remark}

\newtheorem{assumption}{Assumption}

\begin{document}

\def\spacingset#1{\renewcommand{\baselinestretch}%
{#1}\small\normalsize} \spacingset{1}

\if1\anon
{
  \title{\bf On a fast consistent selection of nested models with possibly unnormalized probability densities}
    \author[1,2]{Rong Bian}
    \author[3]{Kung-Sik Chan}
    \author[1,4,5]{Bing Cheng}
    \author[6,7,8]{Howell Tong}
\affil[1]{\small Academy of Mathematics and Systems Science, Chinese Academy of Sciences, Beijing, China}
\affil[2]{\small School of Mathematical Sciences, University of Chinese Academy of Sciences, Beijing, China}
\affil[3]{\small Department of Statistics and Actuarial Science, University of Iowa, Iowa City, USA}  
\affil[4]{\small AMSS Center for Forecasting Science, Chinese Academy of Sciences, Beijing, China}
\affil[5]{\small State Key Laboratory of Mathematical Science, Academy of Mathematics and Systems Science, Chinese Academy of Sciences, Beijing, China}    
\affil[6]{\small Paula and Gregory Chow Institute for Studies in Economics, Xiamen University, Xiamen, China}
\affil[7]{\small Department of Statistics and Data Science, Tsinghua University, Beijing, China}
\affil[8]{\small Department of Statistics, London School of Economics and Political Science, London, UK}
  \maketitle
} \fi

\if0\anon
{
  \bigskip
  \bigskip
  \bigskip
  \begin{center}
    {\LARGE\bf On a fast consistent selection of nested models with possibly unnormalised probability densities}
\end{center}
  \medskip
} \fi

\bigskip
\begin{abstract}
Models with unnormalized probability density functions are ubiquitous in statistics, artificial intelligence and many other fields. However, they  face significant challenges in model selection if the normalizing constants are  intractable. 
Existing methods to address this issue often incur high computational costs, either due to numerical approximations of normalizing constants or evaluation of bias corrections in information criteria. In this paper, we propose a novel and fast selection criterion, MIC, for nested models of possibly dependent data,  allowing direct data sampling from a possibly unnormalized probability density function. 
With a suitable multiplying factor depending only on the sample size and the model complexity, MIC gives a consistent selection  under mild regularity conditions and is computationally efficient. 
Extensive simulation studies and real-data applications demonstrate the efficacy of MIC in the selection of nested models with unnormalized probability densities.
\end{abstract}

\noindent%
\vfill

\newpage
\spacingset{1.8} 

\section{Introduction}
\label{sec:intro}

Models with unnormalized probability density functions are ubiquitous. In statistics, artificial intelligence, statistical mechanics and many other fields, often we only want or are able to stipulate the general shape of the models' distributions without requiring the underlying probability density functions (PDFs) to integrate to unity, hence the notion of  unnormalized PDFs. Here, the normalizing constants are either difficult or impossible to compute explicitly. They arise in various circumstances, e.g. non-conjugacy in Bayesian posteriors \citep{andrade2017exact}, partition functions in statistical mechanics \citep{Frigg_Werndl_2024}, directional distributions of data on a sphere \citep{pewsey2021recent}, Ising model in spatial statistics \citep{friel2013evidence}, and distributions with high-dimensional latent variables \citep{murray2008evaluating}. In short, unnormalized PDFs pose a significant challenge in likelihood-based model comparison.
Addressing it, we focus on the selection of nested models with possibly unnormalized PDFs.

Model selection involves choosing the best statistical model from several candidates based on the observed data, with each candidate having possibly a different number of parameters \citep{rao2001model}. Among the forerunners, 
\cite{akaike1974new} proposed the Akaike Information Criterion (AIC) based on the Kullback-Leibler divergence.
As  the other, \cite{schwarz1978estimating} developed the Bayesian Information Criterion (BIC) via a Laplace approximation of the Bayes factor. 
Unfortunately, none of them is applicable for the selection of  models with unnormalized PDFs.

There are  two main approaches to overcome the challenge. One is to first estimate the intractable normalizing constant by numerical approximations or Markov Chain Monte Carlo. See, for example, \citep{baker2022useful} and
\citep{mackay2003information,congdon2006bayesian}. While simple and intuitive, this approach is computationally intensive, especially for random vectors in high-dimensional settings. Another is to avoid the calculation of the normalizing constant altogether via the score matching method \citep{hyvarinen2005estimation} that provides significant potential 
through efficient computation without the normalizing constant. This method has been applied to Bayesian model selection \citep{dawid2015bayesian, shao2019bayesian} that focuses mainly on the issue of improper priors while the data are still sampled from a normalized distribution. In this approach,  consistent selection is only proved for non-nested models \citep[p.1826]{shao2019bayesian},  with the situation for nested models, such as  polynomial regression models  or autoregressive (AR) models, remaining unclear. 
 
Direct sampling from an unnormalized PDF in model selection is challenging. For this, the following approaches are available. 
Recently, for {\it independent} data, \cite{matsuda2021information} proposed an information criterion
for selection of models with unnormalized PDFs estimated via  noise contrastive estimation (NCE) or score matching. Slightly later, for both independent and {\it Markov} dependent data,  \cite{cheng2024foundation} 
proposed an information criterion for model selection.
In the above information criteria, bias correction is involved and the estimation of the bias term can be computationally intensive. 
For example, with parameter dimension $h$ and sample size $n$, matrix calculations for the bias  in \cite{cheng2024foundation} run in $\mathcal{O}(nh^2 + h^3)$ time. Similar order applies to \cite{matsuda2021information}. 
Besides the computational burden, a more critical limitation of these methods is the absence of model selection consistency.

In this paper, we propose 
a fast information criterion for consistent  selection of nested models of possibly Markov dependent data with unnormalized PDFs. We name the criterion MIC. 
In it, we introduce a multiplying factor $C(n,k)$, which depends only on the sample size $n$ and the order $k$ of the candidate model $M_k$, resulting in a drastic reduction of computation to $\mathcal{O}(1)$. We show that MIC achieves consistent model selection for strictly nested models under mild regularity conditions. Further, we demonstrate the efficacy of MIC  through simulation on AR and polynomial regression models with unnormalized 
PDFs as well as a model with a bivariate von Mises PDF with bounded support. 
Finally, we apply the MIC to real data from diverse domains, including finance, automotive engineering, and wind direction analysis.

This paper is organized as follows. In Section \ref{s2}, we provide a brief review of F-divergence and associated notions. In Section \ref{s3}, we introduce the MIC 
and prove its consistency for model selection under regularity assumptions. 
In Section \ref{s4}, we conduct simulations to demonstrate the efficacy of MIC. In Section \ref{s5}, we apply MIC to various real-world data, offering some insights. In Section \ref{s6}, we conclude with a discussion of our findings and potential directions for future research. Proofs are provided in Appendix \ref{sapp}.

\section{A New Information Criterion}\label{s2}

\subsection{Fisher Divergence}
Let $p(x)$ be a generic PDF 
on $\mathbb{R} ^d$ under the following assumptions: 
\begin{assumption} \label{ass1}
$p(x)$ is twice differentiable on $\mathbb{R} ^d$;
\end{assumption}

\begin{assumption} \label{ass2}
$p(x)$, $\nabla_x p(x)$ and $\nabla_x^2 p(x)$ are all square-integrable on $\mathbb{R} ^d$;
\end{assumption}

\begin{assumption} \label{ass3}
For every $x\in R^d$ with $x=(x_1,\cdots ,x_d)$ and  for each boundary point of $x_i,~i=1,\cdots ,d$, 
		\[
		p(x_1,\cdots,x_{i-1},-\infty ,x_{i+1},\cdots ,x_d)\equiv 0 \text{ and } p(x_1,\cdots,x_{i-1},+\infty ,x_{i+1},\cdots ,x_d)\equiv 0,
		\]
        where, e.g., $p(x_1,\cdots,x_{i-1},\infty ,x_{i+1},\cdots ,x_d)$ denotes  $\lim_{x_i\to\infty} p(x_1,\cdots,x_i,\cdots,x_d)$.
\end{assumption} 
    
Consider the following objective function: 
\begin{align}
   W(x,p)=-||\nabla_x \log p(x)||^2~-~2\Delta_x \log p(x),  \label{w}
\end{align}	
where $\Delta$ denotes the Laplacian, {\it i.e. } $\Delta_x f(x) = \sum_{i=1}^d \frac{\partial^2 f(x)}{\partial^2 x_i}$. 
For PDFs $p$ and $q$ under the above assumptions, the Fisher divergence satisfies the following equation:
\begin{align}\label{defFisherDivergence}
    D_F(p||q)=E_p||\nabla_x \log p(x)~-~\nabla_x \log q(x)||^2
    = E_p[W(x,p)]-E_p[W(x,q)]. 
\end{align}		
 See, e.g., \cite{cheng2024foundation}. 
 Note that $E_p[W(x,p)]$, interpreted as an entropy, is $trace(G_p),$ where $G_p$ is a matrix whose $(i,j)$-th element is $E_p[f_i(x)f_j(x)/p^2(x)],$ $f_i(x) = \partial p/\partial x_i.$ 
See (\cite{chengTong2025discussion}), who introduced the matrix under the name {\it covariate Fisher Information Matrix} based on a Riemannian geometry approach.  

\subsection{GIC Estimate}
Let $\theta$ be the unknown $h$-dimensional parameter vector for $p(x)$, $h\geq 1$. 
When $p$ is the true data PDF and $q$ the model PDF, by equation (\ref{defFisherDivergence}), given PDF $p$ minimizing $D_F(p||q)$  with respect to $q$ is equivalent to maximizing  $E_p[W(x,q)],$
 inspiring the maximum GIC estimate (MGICE). It is worth noting that the MGICE is equivalent to the score matching method \citep{hyvarinen2005estimation}.
 
Specifically, let $p_x$ be the data PDF, and $p_{M(\theta)}$ be a PDF of model $M$ with unknown $h$-dimensional parameter vector $\theta$ ($h\geq 1$) in parameter space $\Theta$. The true parameter $\theta^*$ can be obtained by 
\begin{align}
\theta^*=\arg \min_{\theta \in \Theta}D_F(p_x||p_{M(\theta)})=\arg \max_{\theta \in \Theta}E_p[W(x,p_{M(\theta)})].
\end{align}
Let $x_1,\cdots ,x_n$ be a sample in $\mathbb{R} ^d$ from the data PDF $p_x$. An unbiased estimate of $E_p[W(x,p_{M(\theta)})]$ is given by
\begin{align}
    GIC(M(\theta))=\frac{1}{n}\sum_{i=1}^n 
	W(x_i,p_{M(\theta)}).  \label{fic}
\end{align}
Hence, the MGICE of $\theta$ is as follows
\begin{align}
\hat{\theta}=\hat{\theta}_n=arg \max_{\theta \in \Theta}\{ GIC(M(\theta))\}. \label{MFICE}
\end{align}

\subsection{GIC for Model Selection}
Consider a collection of candidate parametric models $M_1,\cdots,M_K$, denoted as $M_k(\theta_k), k=1,\cdots,K$, with $\theta_k$ being an unknown $h_k$-dimensional parameter vector ($h_k\geq 1$). An unbiased $GIC_c$ for model selection is derived under mild regularity conditions by correcting the bias, $B$, in $n\times GIC$ as follows:
\begin{align}    GIC_c(M_k(\hat{\theta}_k))=GIC_n(M_k(\hat{\theta}_k))-B_k,\label{ficc}
\end{align}
where $\hat{\theta}_k$ is the MGICE of $\theta_k$, $GIC_n(M_k(\hat{\theta}_k))=n\times GIC(M_k(\hat{\theta}_k))$, and 
\begin{align}
    B_k&=-tr\{E_{p_x}[\nabla_{\theta}W(x,p_{M_k(\theta^*)})\nabla_{\theta}^TW(x,p_{M_k(\theta^*)})]E_{p_x}^{-1}[\nabla_{\theta}^2W(x,p_{M_k(\theta^*)})]\}. 
\end{align}
By maximizing $GIC_c$, an appropriate model from $M_1,\cdots, M_K$ is selected. Notably, the $GIC_c$ criterion is essentially equivalent to the Score Matching Information Criterion (SMIC) proposed by \citet{matsuda2021information} up to a sign change.

\section{A New Model Selection Criterion: MIC}\label{s3}

In this section, we propose a fast model selection criterion, MIC, and show the consistency for a finite sequence of strictly nested models under mild regularity conditions.

\subsection{A Fast Model Selection}
Let $x_1,\cdots ,x_n$ be a sample in $\mathbb{R}^d$ from the data PDF $p_x$. Suppose we have a collection of candidate parametric models as described in Section 2.3. 
The $MIC(k)$ of model $M_k$ is defined as
\begin{align}
    MIC(k)=C(n,k)\times GIC(M_k(\hat{\theta}_k)),
\end{align}
where $C(n,k)$ is a constant depending only on $n$ and $k$, $GIC(M_k(\hat{\theta}_k))$ is defined by equation (\ref{fic}), and $\hat{\theta}_k$ is the MGICE of $\theta_k$ for model $M_k$. 
We propose to select the model that maximizes $MIC(k)$. 
The following sections demonstrate the high computational efficiency and the selection consistency of the proposed criterion $MIC$.

\subsection{Computational cost comparison} \label{3.2}
Compared with the bias-corrected criteria GICc \citep{cheng2024foundation}, NCIC1 and NCIC2 \citep{matsuda2021information}, the proposed MIC significantly reduces the computational costs by introducing a factor $C(n,k)$ to bypass the bias correction calculation. Table~\ref{t-cost} compares the computational costs associated with the penalty terms.
Specifically, for GICc, estimating the bias $B$ based on a sample of size $n$ involves a calculation $\mathcal{O}(nh_k^2 + h_k^3)$, which becomes computationally expensive when $h_k$ or $n$ is large. Moreover, NCIC1 and NCIC2 are two versions of NCIC with different computational requirements. Since NCE regards the normalizing constant as an additional parameter and estimates it together with other parameters by generating $\tilde{n}$ noise samples from a known noise distribution, the bias computation for NCIC1 incurs a higher cost of $\mathcal{O}((n+\tilde{n})(h_k+1)^2 + (h_k+1)^3)$. By comparison, the simpler version NCIC2 requires only $\mathcal{O}(n+\tilde{n})$, assuming that the model contains the true distribution. 
However, the use of the multiplying factor $C(n,k)$ in MIC reduces the computational cost even further to $\mathcal{O}(1)$.

\begin{table}[H]
\caption{Computational cost of penalty terms in model selection criteria.} \label{t-cost}
\vspace{10pt}
\resizebox{\textwidth}{!}{
\begin{tabular}{lllll}
\toprule
Criteria        &GICc &NCIC1 &NCIC2 &MIC \\
\midrule
Computational cost  &$\mathcal{O}(nh_k^2 + h_k^3)$&$\mathcal{O}((n+\tilde{n})(h_k+1)^2 + (h_k+1)^3)$ &$\mathcal{O}(n+\tilde{n})$ & $\mathcal{O}(1)$             \\ 
\bottomrule
\end{tabular}
}
\end{table}

\subsection{Consistency of MIC}
The following assumptions are from \cite{song2020sliced} and \cite{cheng2024foundation}. 

\begin{assumption} \label{ass4}
$p_x=p_{M(\theta^*)},$ where $\theta^*$ is the true parameter in $\Theta$. Furthermore, $p_{M(\theta)} \neq p_{M(\theta^*)} $ whenever $\theta \neq \theta^*$.
\end{assumption}

\begin{assumption} \label{ass5}
$p_{M(\theta)}(x) > 0,~\forall \theta \in \Theta$ and $\forall x$.
\end{assumption}

\begin{assumption} \label{ass6}
The parameter space $\Theta$ is compact. 
\end{assumption}

\begin{assumption} \label{ass7}
Both $\nabla_x^2\log p_{M(\theta)}(x)$ and $[\nabla_x\log p_{M(\theta)}(x)][\nabla_x\log p_{M(\theta)}(x)]^T$ are Lipschitz continuous in respect of 
Frobenius norm. Specifically, $\forall \theta_1,\theta_2 \in \Theta$, 
\[
||\nabla_x^2\log p_{M(\theta_1)}(x)-\nabla_x^2\log p_{M(\theta_2)}(x)||_F\leq L_1(x) ||\theta_1-\theta_2||_2, 
\]
and
\[
||[\nabla_x\log p_{M(\theta_1)}(x)][\nabla_x\log p_{M(\theta_1)}(x)]^T-[\nabla_x\log p_{M(\theta_2)}(x)][\nabla_x\log p_{M(\theta_2)}(x)]^T||_F
\]
\[\leq L_2(x) ||\theta_1-\theta_2||_2. 
\]

In addition, $E_{p_x}[L_1^2(x)]<\infty$ and $E_{p_x}[L_2^2(x)]<\infty.$
\end{assumption}

\begin{assumption} \label{ass8}
For $\theta_1$, $\theta_2$ near $\theta^*$, and $\forall i, j$,
\[
||\nabla_{\theta}^2 \partial_i \partial_j \log p_{M(\theta_1)} - \nabla_{\theta}^2 \partial_i \partial_j \log p_{M(\theta_2)}||_F \leq M_{i,j}(x)||\theta_1 - \theta_2||_2,
\]
and
\[
||\nabla_{\theta}^2 \partial_i \log p_{M(\theta_1)} \partial_j \log p_{M(\theta_1)}- \nabla_{\theta}^2 \partial_i \log p_{M(\theta_2)} \partial_j \log p_{M(\theta_2)}||_F \leq N_{i,j}(x)||\theta_1 - \theta_2||_2.
\]
Here, $\partial_i$ refers to  the partial derivative with respect to the component $x_i$ in the random vector $x=(x_1,\cdots ,x_d).$
\end{assumption}

Note that Assumptions \ref{ass4} and \ref{ass6} are standard conditions for proving the consistency of the maximum likelihood estimation (MLE). Assumption \ref{ass5} is also used by \cite{hyvarinen2005estimation}. Assumption \ref{ass7} defines Lipschitz continuity, while Assumption \ref{ass8} describes Lipschitz smoothness for second derivatives. Based on these assumptions, Proposition 6 in \cite{cheng2024foundation} shows the asymptotic normality of the MGICE, as stated in Lemma \ref{asp: MFICE}.
\begin{lemma} \label{asp: MFICE}
Under Assumptions~\ref{ass1}--\ref{ass8} and let $\hat{\theta}_n$ be the MGICE, we have
\begin{equation}
    \sqrt{n}(\hat{\theta}_n-\theta^*)\xrightarrow{dist} N(0,D^{-1}(\theta^*) \Lambda (\theta^*)D^{-T}(\theta^*)),
\end{equation}
where 
\begin{equation}
   D(\theta^*) =-E_{p_x}[\nabla_{\theta}^2W(x,p_{M(\theta^*)})], \label{matrixD}
\end{equation}
\begin{equation}
   \Lambda (\theta^*)=E_{p_x}[\nabla_{\theta}W(x,p_{M(\theta^*)})\nabla_{\theta}^TW(x,p_{M(\theta^*)})].
\end{equation}
\end{lemma}

Moreover, Table \ref{t-MLE-MGICE} provides a detailed comparison between MLE and MGICE.

\begin{table}[H]
\caption{Comparison between MLE and MGICE.} \label{t-MLE-MGICE}
\vspace{10pt}
\resizebox{\textwidth}{!}{
\begin{tabular}{lll}
\toprule
Aspects        &MLE &MGICE  \\
\midrule
Objective function              & $\log[p_{M(\theta)}(x_i)]$         & $W(x_i,p_{M(\theta)})=-||\nabla_x \log p_{M(\theta)}(x_i)||^2  ~-~2\Delta_x \log p_{M(\theta)}(x_i)$          \\
\hline
Sample function                 & $l_n(M(\theta))=\sum_{i=1}^n \log[p_{M(\theta)}(x_i)]$          & $GIC_n(M(\theta))=\sum_{i=1}^n W(x_i,p_{M(\theta)})$    \\    
\hline
Estimate                 & $\hat{\theta}_{MLE}=arg \max_{\theta \in \Theta}\{ l_n(M(\theta))\}$          & $\hat{\theta}_{GIC}=arg \max_{\theta \in \Theta}\{ GIC_n(M(\theta))\}$    \\    
\hline
Information 1                 & $I (\theta^*)=E_{p_x}[\nabla_{\theta}\log[p_{M(\theta^*)}(x)]\nabla_{\theta}^T\log[p_{M(\theta^*)}(x)]]$          & $ \Lambda (\theta^*)=E_{p_x}[\nabla_{\theta}W(x,p_{M(\theta^*)})\nabla_{\theta}^TW(x,p_{M(\theta^*)})]$    \\ 
\hline
Information 2               & $J(\theta^*) =-E_{p_x}[\nabla_{\theta}^2 \log[p_{M(\theta^*)}(x)]]$          & $ D(\theta^*) =-E_{p_x}[\nabla_{\theta}^2W(x,p_{M(\theta^*)})]$    \\ 
\hline
Consistency            &$\hat{\theta}_{MLE} \stackrel{p}{\longrightarrow} \theta^*~\text{ as }n\longrightarrow \infty$ &$\hat{\theta}_{GIC} \stackrel{p}{\longrightarrow} \theta^*~\text{ as }n\longrightarrow \infty$\\
\hline
Asymptotics              & $\sqrt{n}(\hat{\theta}_{MLE}-\theta^*)\xrightarrow{dist} N(0,J^{-1}(\theta^*) I (\theta^*)J^{-T}(\theta^*))$          & $ \sqrt{n}(\hat{\theta}_{GIC}-\theta^*)\xrightarrow{dist} N(0,D^{-1}(\theta^*) \Lambda (\theta^*)D^{-T}(\theta^*))$    \\ 
\bottomrule
\end{tabular}
}
\end{table}

Next, we claim that under suitable regularity conditions, $n\times [\log GIC(M_k(\hat{\theta}_k)) - \log GIC(M_{k_0}(\hat{\theta}_{k_0}))]$ converges weakly to some non-negative distribution, where $k_0$ is the smallest $k$ such that $M_k$ contains the true model. 

\begin{proposition} \label{prop1}
Let $M_1,M_2,\cdots,M_K$ be a finite sequence of strictly nested models, i.e., $M_k \subsetneq M_{k+1}$ for all $k=1,\cdots,K-1$, with $k_0$ being the smallest $1\le k\le K$ such that $M_k$ contains the true model. 
Let $M_k(\theta_k)$ denote the model $M_k$, where $\theta_k$ is an unknown $h_k$-dimensional parameter vector ($h_k \geq 1$).
Consider the model $M = M_k$ for some $k > k_0$ with its parameter $\theta$ partitioned into two sub-vectors $\alpha$ and $\beta$ such that the true parameter $\theta^*$ obtains when $\alpha = \alpha^*$, $\beta =\beta^* = 0$. Similarly, partition $D(\theta^*)$ defined in (\ref{matrixD}) into a 2 by 2 block matrix:
\begin{align}
D(\theta^*)=
\begin{pmatrix}
 D(\alpha^*,\alpha^*) & D(\alpha^*,\beta^*)\\
 D(\beta^*,\alpha^*)  & D(\beta^*,\beta^*)
\end{pmatrix}.  
\end{align}
Assume that $M_{k_0}$ is obtained from $M$ by constraining $\beta = 0$. Under Assumptions~\ref{ass1}--\ref{ass8}, then 
\begin{align}
    n\times [\log GIC(M_k(\hat{\theta}_k)) - \log GIC(M_{k_0}(\hat{\theta}_{k_0}))] \xrightarrow{dist}  Z^TA(\theta^*)Z,
\end{align} and
\begin{align}
    A(\theta^*)=\frac{1}{2H_G(p^*)}B^T(\theta^*)\{D(\beta^*,\beta^*)-D(\beta^*,\alpha^*)D^{-1}(\alpha^*,\alpha^*)D(\alpha^*,\beta^*)\}B(\theta^*),
\end{align}
\begin{align}
    B(\theta^*)=\{D(\beta^*,\beta^*)-D(\beta^*,\alpha^*)D^{-1}(\alpha^*,\alpha^*)D(\alpha^*,\beta^*)\}^{-1} \nonumber\\
    \times
\begin{pmatrix}
 -D(\beta^*,\alpha^*)D^{-1}(\alpha^*,\alpha^*) & I
\end{pmatrix}
\Lambda^{1/2} (\theta^*),
\end{align}    
where $Z$ is a $h_k$-dimensional standard normal random vector, $p^*=p_{M(\alpha^*,\beta^*)}$, $H_G(p^*)=E_{p^*}||\nabla_x \log p^*(x)||^2$, and $I$ denotes the identity matrix of order $(h_k - h_{k_0})$.
Furthermore, $Z^T A(\theta^*) Z$ is a non-negative random variable.
\end{proposition}

Now, we establish the consistency of our MIC method.
\begin{theorem}\label{thm-TFIC-consistency}
Let $M_1,M_2,\cdots,M_K$ be a finite sequence of strictly nested models, with $k_0$ being the smallest $1\le k\le K$ such that $M_k$ contains the true model. Under Assumptions~\ref{ass1}--\ref{ass3}, if the following conditions are satisfied:
\begin{enumerate}
\renewcommand{\labelenumi}{(\theenumi)}
    \item For any $K\ge k\ge k_0$, Assumptions~\ref{ass4}--\ref{ass8} hold and the MGICE of their parameter is $\sqrt{n}$-consistent with an asymptotic normal distribution whose covariance matrix is invertible.
    \item For any $1\le k< k_0$, Assumptions~\ref{ass6}--\ref{ass7} hold.
    \item $C(n,k)$ is such that (\romannumeral 1) for any $k$, $C(n,k)\to 1$ as $n\to \infty$ and (\romannumeral 2) for any $K\ge k_1>k_2\ge 1$, $n\times \log\{C(n,k_1)/C(n,k_2)\}\to -\infty$ as $n\to \infty$. \label{item3}
\end{enumerate}
Let $ \hat{k}=\arg \max _{1\le k\le K}MIC(k)$. Then $ \hat{k}$ converges to $k_0$, in probability.
\end{theorem}

\begin{remark}
For ease of exposition, we have, so far, assumed that the data are  independent and identically distributed (IID).  However,  Proposition~\ref{prop1} and Theorem~\ref{thm-TFIC-consistency} can be extended to dependent data, under suitable regularity conditions. For instance,  for  stationary ergodic finite-order homogeneous Markov processes including autoregressive models, we can generalize GIC  as follows, for an order-$k$ Markov process:
\begin{align}
    CGIC(M(\theta))=\frac{1}{n-k}\sum_{i=k+1}^n 
	W(x_i,p_{M(\theta)}(\cdot|x_{i-1},\cdots, x_{i-k})),  \label{cfic}
\end{align}
where $p_{M(\theta)}(\cdot|x_{i-1},\cdots, x_{i-k})$ is the conditional PDF of $x_i$ given its lags 1 to $k$. It follows from Assumptionss~\ref{ass1}--\ref{ass3} that $\nabla_\theta CGIC(M(\theta))$, evaluated at the true parameter, is a martingale difference sequence. By appealing to some variant of the martingale central limit theorem \citep[Chapter 3]{hall2014martingale}, the proofs of Proposition~\ref{prop1} and Theorem~\ref{thm-TFIC-consistency} for the IID case can be easily extended to the case of stationary ergodic finite-order Markov processes, with MGICE and MIC modified accordingly.  
\end{remark}

Let us consider the following choices of $C(n, k)$:
(1) $\exp\{-2\#(M_k)/n\}$ and (2) $n^{-\#(M_k)/n}$, 
where $\#(M_k)$ is the number of independently adjusted parameters in model $M_k$.
From Theorem \ref{thm-TFIC-consistency}, the following corollaries follow immediately.
\begin{corollary}
Let $M_1,M_2,\cdots,M_K$ be a finite sequence of strictly nested models, with $k_0$ being the smallest $1\le k\le K$ such that $M_k$ contains the true model. If $K=k_0$, under Assumptions~\ref{ass1}--\ref{ass8}, then $MIC1=\exp\{-2\#(M_k)/n\}\times GIC(M_k)$ will consistently select the true model $M_{k_0}$.
\end{corollary}
Note that MIC1 may fail to provide consistent selection when $K>k_0$.
\begin{corollary}
Let $M_1,M_2,\cdots,M_K$ be a finite sequence of strictly nested models, with $k_0$ being the smallest $1\le k\le K$ such that $M_k$ contains the true model. Under Assumptions~\ref{ass1}--\ref{ass8}, then $MIC2=n^{-\#(M_k)/n}\times GIC(M_k)$ will consistently select the true model $M_{k_0}$.
\end{corollary}
 
In the following sections, we will compare the performance of MIC1 and MIC2 through simulations and applications.
		
\section{Simulation Study}\label{s4}
In this section, we assess the efficacy of MIC1 and MIC2 for models with unnormalized PDFs, in comparison with the three bias-corrected criteria GICc, NCIC1, and NCIC2. 
In Section \ref{4.1}, we study the consistency of MGICE for Baker PDF. In Sections \ref{4.2} and \ref{4.3}, we evaluate the consistency (or otherwise) of MIC1 and MIC2 and the MGICE for two true models, namely the AR model and the polynomial regression model, each defined on $\mathbb{R}$ and driven by Baker noise/errors. In Section \ref{4.4}, we examine a model with a bvariate von Mises PDF on bounded support. Upon replications, the empirical distribution of the orders selected by MIC is obtained, as well as the average MGICE and its standard deviation (SD) for the model parameters under the true order. 

\subsection {Baker PDF} \label{4.1}
First, we evaluate the efficacy of MGICE for unnormalized PDFs. 
Consider the Baker PDF \citep{baker2022useful}, $N \times t (\alpha,k)$, defined as the product of a normal PDF and a Student t-PDF:
\begin{align}
f_{N \times t}{(x\mid \alpha,k)}=C(\theta)\frac{\exp\{-\frac{\alpha x^2}{2}\}}{(1+x^2)^k}, \label{def-Nt}
\end{align}
for $-\infty<x<\infty$, where $\alpha>0$, $k>0$, $\theta=(\alpha,k)^T$ and $C(\theta)$ is the normalizing constant. 
Baker PDF is fat-tailed, with the parameters $\alpha$ and $k$ collectively controlling the tail behavior.
Specifically, $\alpha$ represents the scale parameter, and $k$ is the power parameter, which is related to the degrees of freedom $\nu$ of a t-PDF  via $k = (\nu+1)/2$.

The normalizing constant $C(\theta)$ is generally computationally intractable, except in certain special cases, such as when $k$ is an integer. Prior to the MLE, \cite{baker2022useful} either (i) derived $C(\theta)$ under the above condition, or (ii) used numerical quadrature to approximate $C(\theta)$ for general $k$. Instead of the MLE, we employ MGICE for data fitting, which circumvents the need for normalizing constants. 

Considering the affine transformation $X=(Y-\mu)/s$ in \cite{baker2022useful}, the PDF of $Y$ is $\frac{1}{s}f_{N \times t}(\frac{y-\mu}{s}|\alpha,k)$, with 4 parameters $(\mu,s,\alpha,k)$, where $-\infty<\mu<\infty$ and $s>0$. Let $\{y_i, i = 1, 2, \cdots, n\}$ represent $n$ observations and denote $f_{y_i}(\theta) = \frac{1}{s}f_{N \times t}(\frac{y_i-\mu}{s}|\alpha,k)$. Following routine derivations, we have
\begin{align}
    \nabla_{y_i} \log(f_{y_i})&=-\frac{1}{s}[\alpha x_i+\frac{2kx_i}{1+x_i^2}],
    \label{Nt1}
\end{align}
and
\begin{align}
    \Delta_{y_i} \log(f_{y_i})&=-\frac{1}{s^2}[\alpha+\frac{2k(1-x_i^2)}{(1+x_i^2)^2}],
    \label{Nt2}
\end{align}
where $x_i=(y_i-\mu)/s$.
From equations (\ref{w}), (\ref{fic}) and (\ref{MFICE}), we obtain the MGICE, $\hat\theta_{GIC}$. 

We conduct 100 replications to evaluate the performance of the MGICE, with sample size $n= 1000,3000, \text{and }5000$, where the true parameter values are set to $(\mu^*,s^*,\alpha^*,k^*)=(0.3,0.5,0.5,1.5)$. Samples are generated using the acceptance-rejection method, by generating random numbers from the $N(0,1/\alpha)$ distribution. For numerical optimization in MGICE, we use the Adaptive Moment Estimation (Adam) algorithm \citep{2015-kingma} to jointly optimize all parameters. Similar to \cite{baker2022useful}, for all experiments involving the Baker PDF, we use the sample mean and standard deviation as starting values for $\mu$ and $s$, and regular initial values for $\alpha$ and $k$, e.g. $\alpha =0.25$ or 1, and $k=1$ or 2. 
Table \ref{ts0} presents the average MGICE and its SD, showing good overall consistency of MGICE, although there is still room for improvement for the parameters $s,\alpha,$ and $k$.  

\begin{table}[htbp]
\small
\caption{The average MGICE and its SD for Baker PDF.}\label{ts0}
\begin{center}
\begin{tabular}{lllll}
\toprule
\multirow{2}{*}{$N$} & \multicolumn{4}{c}{Parameter MGICE (SD)} \\ 
\cline{2-5} 
                   & $\mu$     &$s$ & $\alpha$     & $k$        \\
\midrule
1000               & 0.30 (0.02) &0.52 (0.10) & 0.52 (0.25) & 1.62 (0.39)  \\
3000               & 0.30 (0.01) &0.53 (0.06) & 0.48 (0.14)  & 1.68 (0.19) \\
5000               & 0.30 (0.01) &0.53 (0.04) & 0.48 (0.10)  & 1.70 (0.14)  \\
\bottomrule
\end{tabular}
\end{center}
\end{table}

\subsection{AR Model with Baker Noise} \label{4.2}

Now, we consider an AR model with Baker distributed noise and use MIC method to select the model order. The stationary mean-centered AR model of order p with Baker noise is given by
\begin{align}
    X_t-c=a_1(X_{t-1}-c)+\cdots+a_p(X_{t-p}-c)+s\varepsilon_t,  \label{e_AR}
\end{align}
where $-\infty<a_1,\cdots,a_p,c<\infty$, $s>0$, and $\varepsilon_t \sim N \times t (\alpha,k),$ identically and independently.  

Let $\{x_t, t=1,2, \cdots, N\}$ denote $N$ observations from the above model and denote the parameter as $\theta=(a_1,\cdots,a_p,c,s,\alpha,k)^T$. Let $y_t=x_t-c$ and $\mu_t=a_1 y_{t-1}+\cdots+a_p y_{t-p}$. Since $\varepsilon_t=(y_t-\mu_t)/s$, the conditional density of $X_t$ is $f_{x_t}=f_p(x_t\mid x_{t-1},\cdots,x_{t-p},\theta)=\frac{1}{s}f_{N \times t}(\frac{y_t-\mu_t}{s}|\alpha,k)$. From equations (\ref{Nt1}) and (\ref{Nt2}), we have $\nabla_{x_{t}} \log(f_{x_t})$ and $\Delta_{x_{t}} \log(f_{x_t})$.

Suppose we have a collection of candidate AR(p) models for $p = 1,\cdots,L$, $L$ being the maximum possible order. Then, $GIC_N$ is given by
\begin{align}
    GIC_N(\theta)=\sum_{t=1}^L W(x_t,f_{x_t}(\theta)) + \sum_{t=L+1}^N W(x_t,f_{x_t}(\theta)).
\end{align}
where $W(x_t,f_{x_t})=-||\nabla_{x_t} \log (f_{x_t})||^2-2\Delta_{x_t} \log (f_{x_t})$. 

Discarding the first sum because $W(x_t,f_{x_t}(\theta))$ for $t=1, \cdots L$ are unavailable in the AR(L) model and denoting $n = N-L$, we have 
\begin{align}
    GIC(\theta)&=\frac{1}{n}\sum_{t=L+1}^{N} W(x_t,f_{x_t}(\theta)). \label{e_1_3}
\end{align}

We conduct 100 replications to obtain the frequency distribution of the selected model orders, ranging from 1 to 10, using GICc, MIC1 and MIC2, with sample size  $N= 1000, 3000,$ and $5000$. The true model order $p^*$ is set to 3, with true parameters $(a_1^*,a_2^*,a_3^*,c^*,s^*,\alpha^*,k^*)$ specified at $(0.50,-0.25,0.10,3.00,0.50,0.50,1.50)$. 
Here, NCIC1 and NCIC2 are not considered in this experiment, as they are developed under the assumption of independent and identically distributed observations, and their applicability to dependent data in stationary AR models remains unclear.
In each replication, the first 200 data are discarded to ensure stationarity.

Table \ref{ts3} shows that MIC2 tends to underestimate the model order for smaller $N$ but increasingly selects the correct order as $N$ grows, outperforming the other criteria overall.
The underfitting for small $N$ is not unexpected due to the small value of $a_3^*$.  In contrast, MIC1 overestimates the model order with high probability even for large $N$. Furthermore, Table \ref{ts4} reports the average MGICE and its SD for the parameters in the true model, suggesting strong consistency, especially for the AR coefficients and the mean-centering parameter $c$,  when $N$ is large.

\begin{table}[htbp]
\footnotesize
\caption{The frequency distribution of selected orders for the AR(p) model.}\label{ts3}
\begin{center}
\begin{tabular}{llllllllllllll}
\toprule
\multirow{2}{*}{$N$} & & \multirow{2}{*}{Method} & &\multicolumn{10}{c}{Selected model order $p$} \\ 
\cline{5-14} 
& &   &  & 1  & 2  & 3  & 4 & 5 & 6 & 7 &8 & 9 & 10 \\
\midrule
\multirow{3}{*}{1000} 
& &\underline{GICc} & & 0          & 13          & \underline{57} & 18         & 6          & 1          & 3          & 2          & 0          & 0           \\
& &MIC1 & & 0          & 3          & 22 & 13         & 11          & 1          & 6         & 13          & 14          & 17          \\
& &MIC2 & & 1          & 12 &52          & 9          & 7          & 0          & 4          & 2          & 6          & 7           \\
\hline

\multirow{3}{*}{3000} 
& &GICc & & 0          & 0 &71 & 16          & 2          & 4          & 3          & 2          & 1          & 1           \\
& &MIC1 & & 0          & 0          &28 & 6          & 7         & 9          & 9          & 5          & 16          & 20           \\
& &\underline{MIC2} & & 0          & 1 &\underline{81} & 11          & 0          & 4          & 1          & 0          & 0          & 2           \\
\hline
\multirow{3}{*}{5000} 
& &GICc & & 0          & 0 & 74 & 16          & 3          & 2          & 2          & 2          & 0          & 1           \\
& &MIC1 & & 0          & 0           & 29 & 10         & 8          & 7          & 9          & 10          & 7         & 20          \\
& &\underline{MIC2} & & 0          & 0          & \underline{88} & 5          & 2          & 1          & 2          & 1          & 0          & 1         \\
\bottomrule
\end{tabular}
\end{center}
\end{table}

\begin{table}[htbp]
\caption{The average MGICE and its SD for the AR(3) model with Baker noise.}\label{ts4}
\vspace{10pt}
\resizebox{\textwidth}{!}{
\begin{tabular}{llllllll}
\toprule
\multirow{2}{*}{$N$} & \multicolumn{7}{c}{Parameter MGICE (SD)} \\ 
\cline{2-8} 
                  & $a_1$ &$a_2$ &$a_3$                 &$c$ &$s$ & $\alpha$ &$k$    \\
\midrule
1000               & 0.51 (0.13) &-0.24 (0.11) &0.11 (0.13) & 3.01 (0.11)& 0.51 (0.14)& 0.43 (0.26)& 1.56 (0.61)  \\
3000               &0.50 (0.03)  &-0.26 (0.03)  &0.10 (0.02) & 3.00 (0.01)& 0.51 (0.10)& 0.45 (0.14)& 1.61 (0.42)   \\
5000              & 0.50 (0.02) &-0.25 (0.02)  &0.10 (0.02) & 3.00 (0.01)& 0.52 (0.07)& 0.46 (0.10)& 1.67 (0.31)    \\
\bottomrule
\end{tabular}
}
\end{table}

\subsection{Polynomial Regression Model with Baker Distributed Errors} \label{4.3}

Next, we consider a polynomial regression model with Baker distributed errors and use MIC  to select the model. The polynomial regression model of degree p 
is given by
\begin{align}
    y=\beta_1x+\cdots+\beta_px^p+c+s\varepsilon,\label{e_1_1}
\end{align}
where $-\infty<\beta_1,\cdots,\beta_p,c<\infty$, $s>0$, and $\varepsilon \sim N \times t (\alpha,k),$ identically and independently. 

Let $\{(x_i,y_i), i=1,2, \cdots, n\}$ denote $n$ observations from the above model and denote the parameter as $\theta=(\beta_1,\cdots,\beta_p,c,s,\alpha,k)^T$. Let $\mu_i=\beta_1x_i+\cdots+\beta_px_i^p+c$ and $\varepsilon_i=(y_i-\mu_i)/s$, 
The conditional density of $y_i$ is given by 
$f_{y_i}=f_p(y_i\mid x_i,\theta)=\frac{1}{s}f_{N \times t}(\frac{y_i-\mu_i}{s}|\alpha,k).$ From equations (\ref{Nt1}) and (\ref{Nt2}), we have $\nabla_{y_{i}} \log(f_{y_i})$ and $\Delta_{y_{i}} \log(f_{y_i})$.

Suppose we have a collection of candidate models of degree $p$, for $p = 1,\cdots,L$, $L$ being the maximum possible degree. For each model, from equations (\ref{w}), (\ref{fic}), and (\ref{MFICE}), $GIC$ and its MGICE are obtained.

We conduct 100 replications to obtain the frequency distribution of the selected model degree $p$, ranging from 1 to 10, using GICc, NCIC1, NCIC2, MIC1 and MIC2, with sample size  $n= 300, 500, 1000, 3000,$ and $5000$. The true model order $p^*$ is set to 3, with true parameters $(\beta_1^*, \beta_2^*, \beta_3^*,c^*,s^*, \alpha^*,k^*)$ specified as $(-1.5,2.0,5.0,3.0,0.5,0.5,1.5)$.
For NCIC1 and NCIC2, noise samples of the same size are generated from the classically estimated polynomial regression model distribution using the least squares method under the Gaussian errors assumption.

\begin{figure}[h]
\begin{center}
\includegraphics[width=0.9\textwidth]{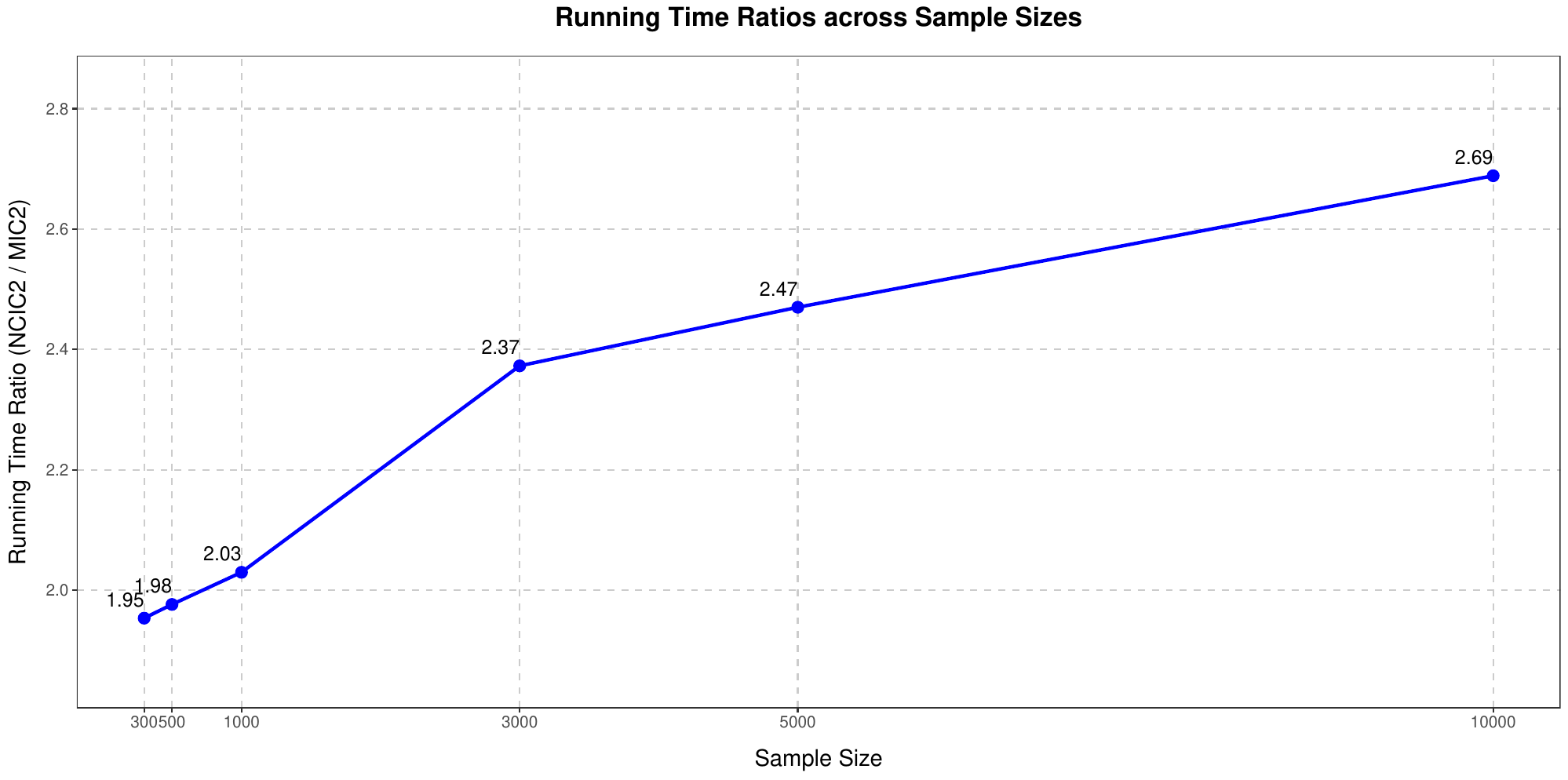}
\end{center}
\caption{Empirical running time ratios between NCIC2 and MIC2 for the polynomial regression model across different sample sizes.}\label{fig_time}
\end{figure}

Table \ref{ts5} suggests that both MIC1 and MIC2 consistently select the correct model degree with increasing sample size. Although MIC2 shows slightly lower accuracy than NCIC2 when the sample size is small, its performance remains satisfactory and becomes comparable to that of NCIC2 as $n$ increases, eventually achieving the best performance among the criteria.
Moreover, MIC2 exhibits substantially higher computational efficiency than NCIC2. 
In addition to the theoretical complexity comparison in Section~\ref{3.2}, we empirically compare the running time ratio of the two methods across different sample sizes, with running time averaged over 10 simulation replications. The experiments were conducted using Python 3.10 on a computer with an Intel Xeon Gold 6133 CPU.
The results, reported in Figure~\ref{fig_time}, indicate that NCIC2 not only requires more running time than MIC2 but also shows a much faster growth in running time as the sample size increases. 
Table \ref{ts6} reports the average MGICE and its SD for the parameters in the true model, showing good consistency as $n$ increases, although at a slightly slower rate for $\alpha$.

\begin{table}[htbp]
\tiny
\caption{The frequency distribution of selected degrees for the polynomial regression model.}\label{ts5}
\begin{center}
\begin{tabular}{llllllllllllll}
\toprule
\multirow{2}{*}{$n$} & & \multirow{2}{*}{Method} & &\multicolumn{10}{c}{Selected model degree $p$} \\ 
\cline{5-14} 
&  & &    & 1  & 2  & 3  & 4 & 5 & 6 & 7 &8 & 9 & 10 \\
\midrule
\multirow{5}{*}{300} 
& &GICc & & 0          & 0          &67 & 23         & 4          & 2          & 3          & 0          & 1          & 0           \\
& &NCIC1 & & 8          & 0 &68 & 10          & 7          & 2          & 2          & 1          & 1          & 1           \\
& &\underline{NCIC2} & & 0          & 0 &\underline{96} & 4          & 0          & 0          & 0          & 0          & 0          & 0           \\
& &MIC1 & & 0          & 0          & 77 & 14          & 6          & 1          & 2          & 0          & 0          & 0           \\
& &MIC2 & & 0          & 0 &81 & 11          & 5          & 1          & 2          & 0          & 0          & 0           \\
\hline
\multirow{5}{*}{500} 
& &GICc & & 0          & 0          & 85 & 9         & 4          & 1          & 1         & 0          & 0          & 0           \\
& &NCIC1 & & 4          & 1 &66 & 15          & 8          & 1          & 0          & 1          & 3          & 1           \\
& &\underline{NCIC2} & & 0          & 0 &\underline{91} & 9          & 0          & 0          & 0          & 0          & 0          & 0           \\
& &MIC1& & 0          & 0 &84 & 9          & 5          & 1          & 1          & 0          & 0          & 0          \\
& &MIC2 & & 0          & 0 &87 & 7          & 4          & 1          & 1          & 0          & 0          & 0           \\
\hline
\multirow{5}{*}{1000} 
& &GICc & & 0          & 0          & 93 & 7         & 0          & 0          & 0          & 0          & 0          & 0           \\
& &NCIC1 & & 1          & 0 &78 & 11          & 3          & 2          & 2          & 0          & 2          & 1           \\
& &NCIC2 & & 0          & 0 &96 & 4          & 0          & 0          & 0          & 0          & 0          & 0           \\
& &MIC1 & & 0          & 0          & 94 & 6          & 0          & 0          & 0          & 0          & 0          & 0           \\
& &\underline{MIC2} & & 0          & 0 &\underline{97} & 3          & 0          & 0          & 0          & 0          & 0          & 0           \\
\hline
\multirow{5}{*}{3000} 
& &GICc & & 0          & 0          & 97 & 3         & 0          & 0          & 0          & 0          & 0          & 0           \\
& &NCIC1 & & 0          & 0          & 77 & 12         & 5          & 2          & 2          & 2          & 0          & 0           \\
& &\underline{NCIC2} & & 0          & 0 &\underline{98} & 2          & 0          & 0          & 0          & 0          & 0          & 0           \\
& &MIC1 & & 0          & 0 &96 & 4          & 0          & 0          & 0          & 0          & 0          & 0          \\
& &MIC2 & & 0          & 0 &96 & 4          & 0          & 0          & 0          & 0          & 0          & 0          \\
\hline
\multirow{5}{*}{5000} 
& &\underline{GICc} & & 0          & 0          &\underline{97} & 3         & 0          & 0          & 0          & 0          & 0          & 0           \\
& &NCIC1 & & 0          & 0 &86 & 12          & 0          & 1          & 0          & 1          & 0          & 0           \\
& &NCIC2 & & 0          & 0 &91 & 9          & 0          & 0          & 0          & 0          & 0          & 0           \\
& &\underline{MIC1} & & 0          & 0 &\underline{97} & 3          & 0          & 0          & 0          & 0          & 0          & 0         \\
& &\underline{MIC2} & & 0          & 0 &\underline{97} & 3          & 0          & 0          & 0          & 0          & 0          & 0       \\
\bottomrule
\end{tabular}
\end{center}
\end{table}

\begin{table}[htbp]
\caption{The average MGICE and its SD for the cubic polynomial regression model.}\label{ts6}
\vspace{10pt}
\resizebox{\textwidth}{!}{
\begin{tabular}{llllllll}
\toprule
\multirow{2}{*}{$n$} & \multicolumn{7}{c}{Parameter MGICE (SD)} \\ 
\cline{2-8} 
                  & $\beta_1$ &$\beta_2$ &$\beta_3$                 &$c$ &$s$ & $\alpha$ &$k$    \\
\midrule
300               & -1.52 (0.47) &2.03 (0.52) &4.95 (0.46) & 2.95 (0.53)& 0.90 (0.65)& 0.74 (2.66)& 3.39 (4.12)  \\
500               & -1.53 (0.34) &2.01 (0.42) &4.94 (0.37) & 2.98 (0.35)& 0.73 (0.49)& 0.25 (0.39)& 2.76 (2.56)   \\
1000               & -1.50 (0.03) &2.00 (0.01) &5.00 (0.01) & 3.00 (0.02)& 0.56 (0.21)& 0.33 (0.20)& 2.08 (1.27)  \\
3000               & -1.50 (0.02) &2.00 (0.01) &5.00 (0.01) & 3.00 (0.01)& 0.50 (0.14)& 0.41 (0.15)& 1.62 (0.78)   \\
5000              & -1.50 (0.01) &2.00 (0.01) &5.00 (0.00) & 3.00 (0.01)& 0.49 (0.12)& 0.44 (0.13)& 1.51 (0.62)    \\
\bottomrule
\end{tabular}
}
\end{table}

\subsection{A Bivariate Model with a Von Mises PDF} \label{4.4}
As a final case, we use MIC  to select the dimension of the parameter space of a model with unnormalized PDF on bounded support. Consider the bivariate von Mises PDF \citep{singh2002probabilistic} of two circular random variables $X_1,X_2$, given by 
\begin{align}
    f(x_1,x_2\mid \theta)=C(\theta) \exp\{&\kappa_1 \cos(x_1-\mu_1)+\kappa_2 \cos(x_2-\mu_2) \nonumber\\ 
    &+\lambda  \sin(x_1-\mu_1) \sin(x_2-\mu_2)\}, \label{e-von}
\end{align}
for $0\le x_1$, $x_2 < 2\pi$, where $\kappa_1$, $\kappa_2 \ge 0$, $0\le \mu_1$, $\mu_2 < 2\pi$, $-\infty <\lambda <\infty$, $\theta=(\kappa_1,\kappa_2,\mu_1,\mu_2,\lambda)^T$ and $C(\theta)$ is the normalizing constant. The parameter $\lambda$ quantifies the dependency between two circular random variables $X_1,X_2$. $C(\theta)$ is computationally intractable, involving an infinite sum of Bessel functions.

Denote $\boldsymbol{x}=(x_{1},x_{2})$ and $f_{\boldsymbol{x}}(\theta)=f(x_1,x_2 \mid \theta)$. 
Let $\{\boldsymbol{x_i}=(x_{i1},x_{i2}), i=1,2, \cdots, n\}$ denote $n$ independent observations from the bivariate model.  
After some routine calculations, we have 
\begin{equation}
    \frac{\partial}{\partial x_{i1}}\log (f_{\boldsymbol{x_i}}) = -\kappa_1 \sin(x_{i1}-\mu_1)+\lambda  \cos(x_{i1}-\mu_1) \sin(x_{i2}-\mu_2),
\end{equation}
\begin{equation}
    \frac{\partial}{\partial x_{i2}}\log (f_{\boldsymbol{x_i}}) = -\kappa_2 \sin(x_{i2}-\mu_2)+\lambda  \sin(x_{i1}-\mu_1) \cos(x_{i2}-\mu_2),
\end{equation}
and
\begin{equation}
    \frac{\partial^2}{\partial x_{i1}^2} \log(f_{\boldsymbol{x_i}}) = -\kappa_1 \cos(x_{i1}-\mu_1)-\lambda  \sin(x_{i1}-\mu_1) \sin(x_{i2}-\mu_2),
\end{equation}
\begin{equation}
    \frac{\partial^2}{\partial x_{i2}^2} \log(f_{\boldsymbol{x_i}}) = -\kappa_2 \cos(x_{i2}-\mu_2)-\lambda  \sin(x_{i1}-\mu_1) \sin(x_{i2}-\mu_2).
\end{equation}

Since random variables $X_1$ and $X_2$ are independent if and only if $\lambda= 0$, there are two candidate models. When $\lambda=0$, the parameter space is 4-dimensional, consisting  of  $(\kappa_1,\kappa_2,\mu_1,\mu_2)$. Denote this model as model $m_1$. For the general
case with non-zero $\lambda \in R$ as defined in equation (\ref{e-von}), the model is denoted as model $m_2$, whose parameter space is 5-dimensional. Clearly, $m_1$ is nested within $m_2$.
For each model, from equations (\ref{w}), (\ref{fic}), and (\ref{MFICE}), $GIC$ and its MGICE are obtained.

We conduct 100 replications to evaluate the performance of model selection between $m_1$ and $m_2$, using GICc, NCIC1, NCIC2, MIC1 and MIC2, with sample size  $n= 300, 500, \text{and }1000$. The true parameter dimension 
is set to 5, with true parameter $\theta^*=(2.0,1.0,1.5,2.5,3.0)^T$. Samples are generated using the acceptance-rejection method, by generating a random number from the uniform distribution on $[0,2 \pi) \times [0,2 \pi)$. 
For NCIC1 and NCIC2, noise samples of the same size are generated from the same uniform distribution.
For numerical optimization in MGICE, we use the Broyden-Fletcher-Goldfarb-Shanno (BFGS) algorithm \citep{wright2006numerical} to jointly optimize all parameters, with all parameters initialized to zero. 
 
\begin{figure}[h]
\begin{center}
\includegraphics[width=0.9\textwidth]{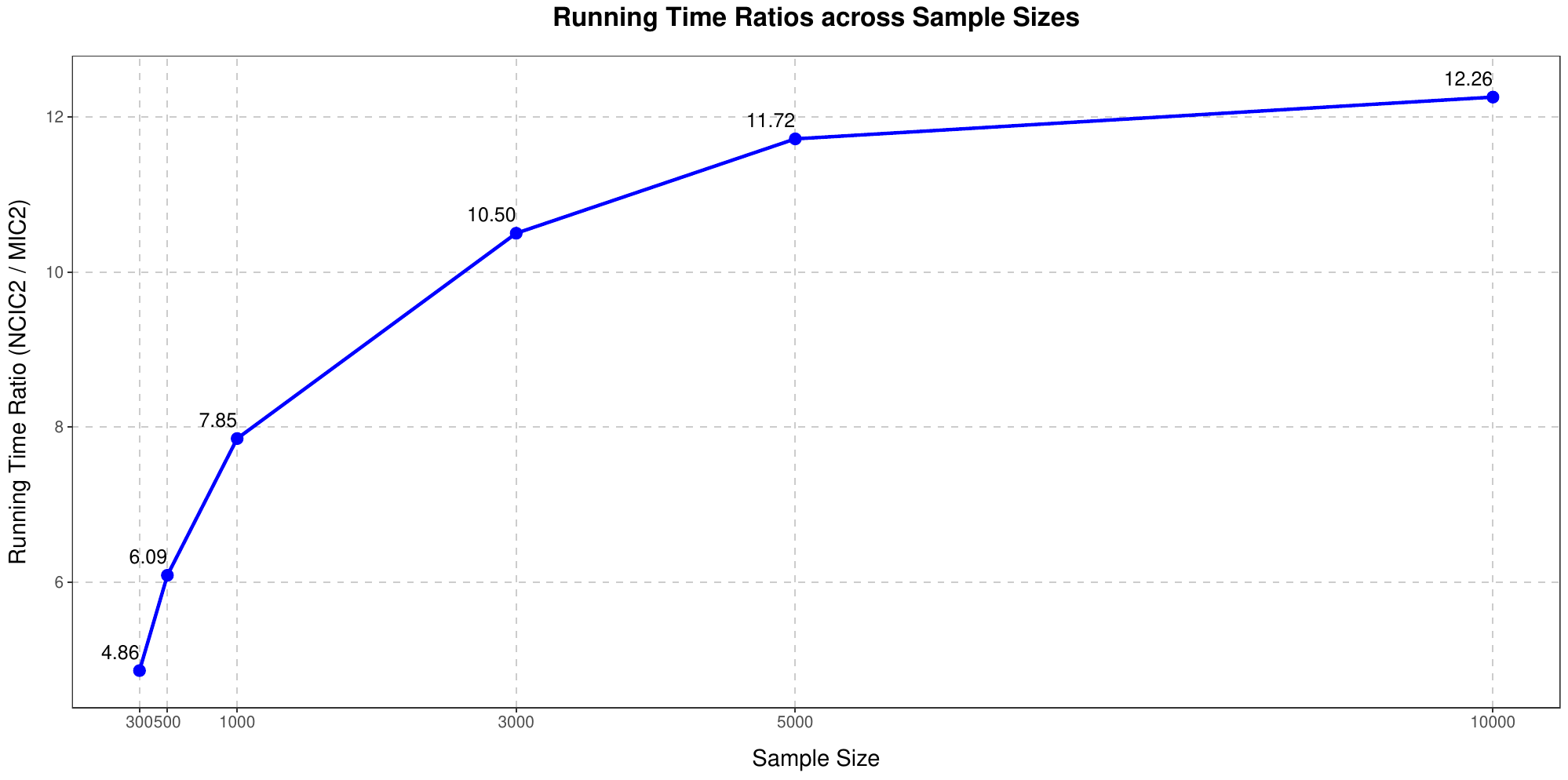}
\end{center}
\caption{Empirical running time ratios between NCIC2 and MIC2 for the bivariate model with a Von Mises PDF across different sample sizes.}\label{fig_time_von}
\end{figure}

In all replications, all criteria correctly select the true parameter dimension 5. However, NCIC2 requires substantially more running time than MIC2, over ten times longer when the sample size exceeds 3000, as shown in Figure ~\ref{fig_time_von}. Table \ref{ts9} presents the average MGICE and its SD for parameters in the true model $m_2$, demonstrating the strong consistency of MGICE especially when $n$ is large.

\begin{table}[htbp]
\small
\caption{The average MGICE and its SD for the model $m_2$.}\label{ts9}
\begin{center}
\begin{tabular}{llllll}
\toprule
\multirow{2}{*}{$n$} & \multicolumn{5}{c}{Parameter MGICE (SD)} \\ 
\cline{2-6} 
                 & $\kappa_1$     & $\kappa_2$     & $\mu_1$    & $\mu_2$      & $\lambda$     \\
\midrule
300               & 2.03 (0.20) & 1.01 (0.14)  & 1.50 (0.05) &2.51 (0.05)   &3.06 (0.24) \\

500                & 2.01 (0.14) & 1.00 (0.11)  & 1.50 (0.04) &2.50 (0.04)  &3.04 (0.17) \\
1000                & 2.01 (0.11) & 1.00 (0.08)  & 1.50 (0.03) &2.50 (0.03)  &3.01 (0.13)\\
\bottomrule
\end{tabular}
\end{center}
\end{table}

\section{Real Data}\label{s5}
In this section, we apply MIC for model selection with real data across three domains. Now, \cite{baker2022useful} has argued that the Baker PDF provides a more realistic framework for modeling real-world data, especially in finance, and serves as an effective tool for assessing  robustness and performing sensitivity analyses. In Sections \ref{5.1} and \ref{5.2}, we consider fitting autoregression to some finance data, and polynomial regression to some car data, with Baker noise/errors, using MIC1 and MIC2 for model selection. We also compare the results  with those based on AIC and BIC with Gaussian noise/errors. 
In Section \ref{5.3}, we consider fitting a  bivariate model with von Mises PDF to some wind direction data.

\subsection{Finance Data} \label{5.1}

\begin{figure}[htbp]
    \centering
    \begin{subfigure}[t]{0.3\textwidth}
        \centering
        \includegraphics[width=\textwidth]{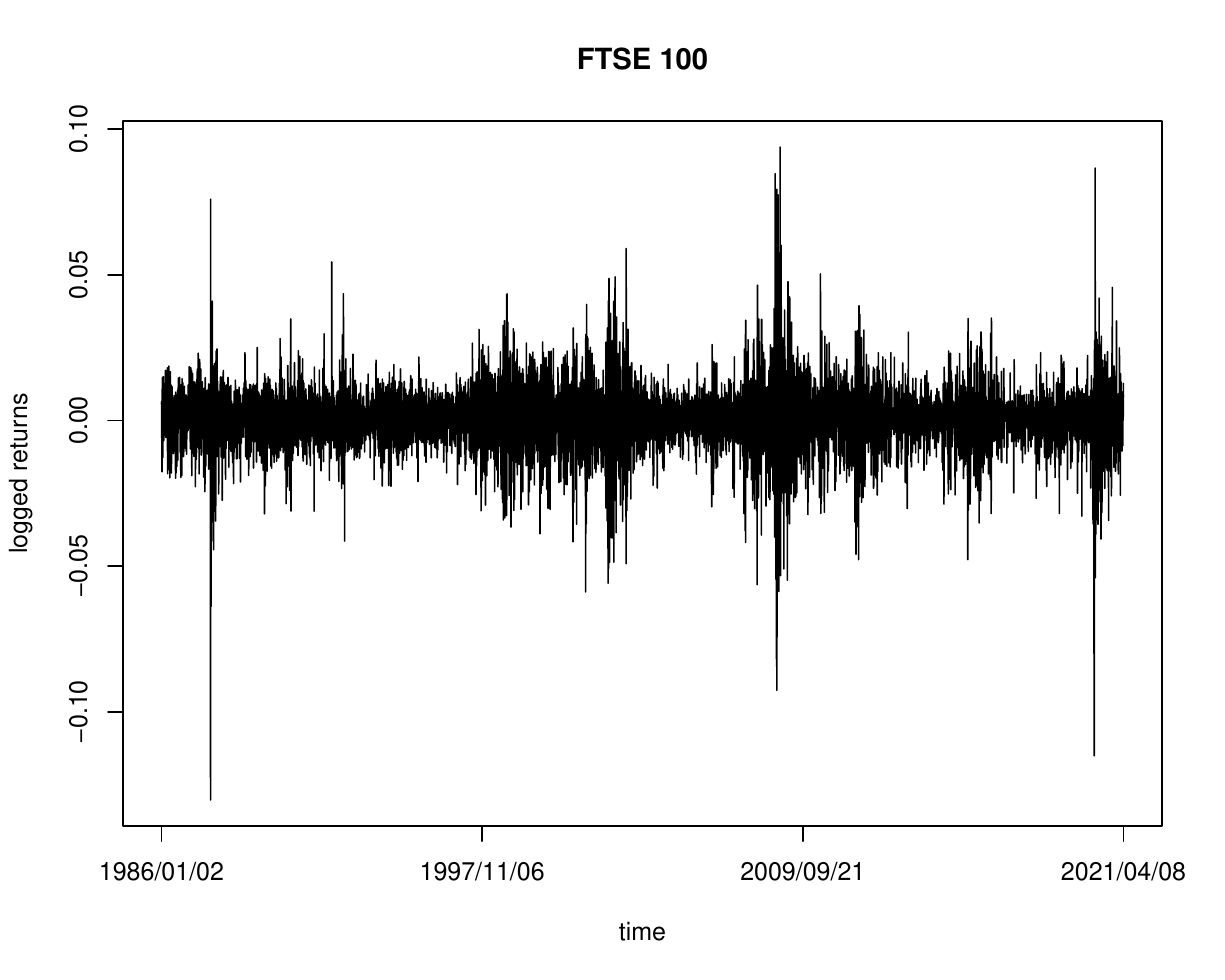}
        \caption{FTSE 100}
    \end{subfigure}
    \begin{subfigure}[t]{0.3\textwidth}
        \centering
        \includegraphics[width=\textwidth]{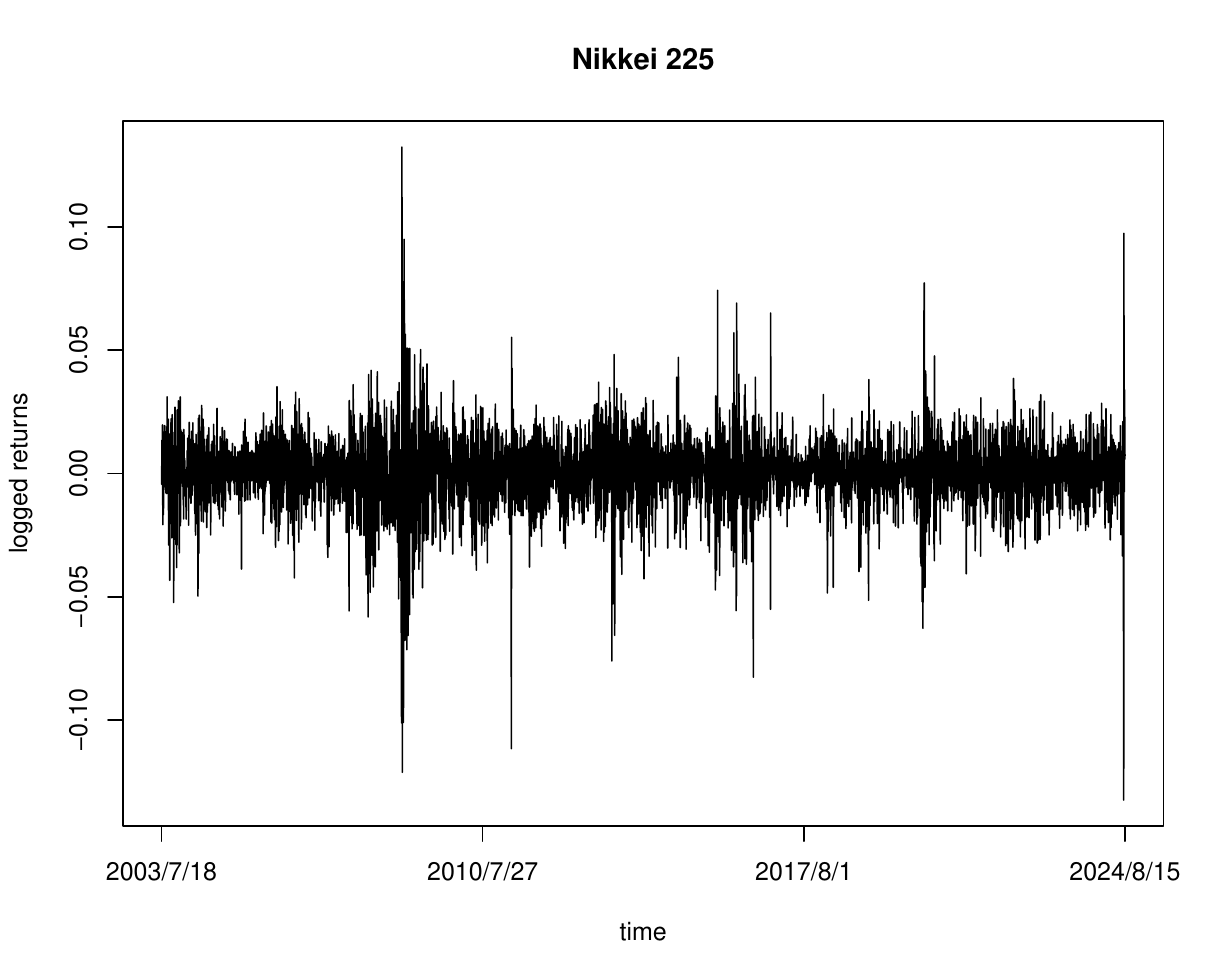}
        \caption{Nikkei 225}
    \end{subfigure}
    \begin{subfigure}[t]{0.3\textwidth}
        \centering
        \includegraphics[width=\textwidth]{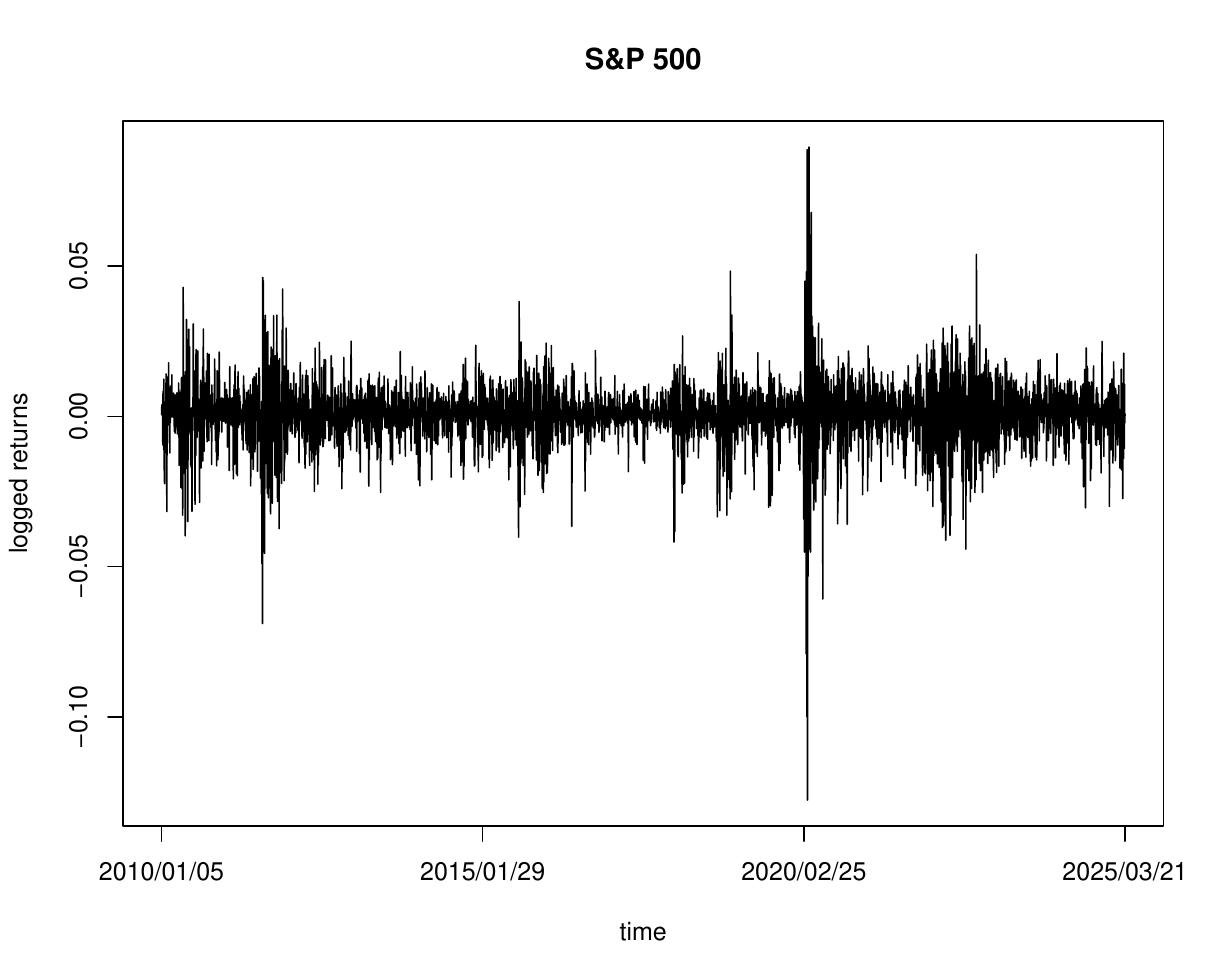}
        \caption{S\&P 500}
    \end{subfigure}
    \caption{Time plots of logged returns on the FTSE 100, Nikkei 225, and S\&P 500.}\label{fig1}
\end{figure}

We analyze the logged returns of three stock market indices: the Financial Times Stock Exchange 100 Index (FTSE 100) with 9013 observations from January 1986 to April 2021, the Nikkei Stock Average (Nikkei 225) with 5159 observations from July 2003 to August 2024, and the Standard \& Poor’s 500 Index (S\&P 500) with 3827 observations from January 2010 to March 2025. Their time plots are displayed in Figure~\ref{fig1}.

\cite{baker2022useful} assumed, for simplicity, independent data and fitted a Baker distribution. Instead, we consider an AR model to allow for data dependence. For comparison, we use MIC1 and MIC2 to select an AR model with Baker noise in equation (\ref{e_AR}), and AIC and BIC to select an AR model with Gaussian noise, given by 
\begin{align}
    X_t-c=a_1(X_{t-1}-c)+\cdots+a_p(X_{t-p}-c)+\varepsilon_t, 
\end{align}
where $\varepsilon_t \sim N(0,\sigma^2)$ independently and identically.
Tables \ref{tr1} and \ref{tr2} summarize the results.
Under Gaussian noise, AIC selects a seventh-order model, while BIC selects a sixth-order model for the FTSE 100. For the Nikkei 225, both AIC and BIC select a first-order model, while for the S\&P 500, both select a ninth-order model.
However, under Baker noise, MIC1 selects a sixth-order AR model, while MIC2 selects a second-order AR model for the FTSE 100, yielding estimates of $\alpha$ and $k$ similar to \cite{baker2022useful}. For the Nikkei 225, MIC1 selects a sixth-order AR model, whereas MIC2 favors a first-order AR model. For the S\&P 500, both MIC1 and MIC2 select a first-order AR model.

\begin{table}[h]
\caption{Selection results for AR model with two noise PDFs.}\label{tr1}
\vspace{10pt}
\resizebox{\textwidth}{!}{
\begin{tabular}{llllllllllll}
\toprule
\multicolumn{12}{c}{$\varepsilon_t \sim N(0,\sigma^2)$}\\
\hline
&$p$   & 1  & 2  & 3  & 4 & 5 &6 &7 &8 &9 &10  \\
\hline
\multirow{2}{*}{FTSE 100}  &AIC & -55538.20 &-55543.89 &-55557.30 &-55571.54 &-55574.02 &-55586.90 &\underline{-55593.47} &-55592.42 &-55591.97 &-55590.44
  \\
&BIC & -55516.88 &-55515.47 &-55521.77 &-55528.91 &-55524.27 &\underline{-55530.05} &-55529.51 &-55521.36 &-55513.80 &-55505.16
    \\
\hline
\multirow{2}{*}{Nikkei 225}  &AIC & \underline{-29127.55} &-29125.80 &-29127.32 &-29126.13 &-29126.43 &-29124.65 &-29122.66 &-29120.81 &-29118.96 &-29118.56
  \\
&BIC & \underline{-29107.90} &-29099.60 &-29094.58 &-29086.84 &-29080.59 &-29072.27 &-29063.72 &-29055.32 &-29046.92 &-29039.98
    \\
\hline
\multirow{2}{*}{S\&P 500}  &AIC & -23789.81 &-23801.46 &-23800.63 &-23808.10 &-23806.12 &-23825.54 &-23854.07 &-23867.60  & \underline{-23879.13} &-23877.23
  \\
&BIC & -23771.06 &-23776.47 &-23769.38 &-23770.60 &-23762.38 &-23775.54 &-23797.82 &-23805.10 &\underline{-23810.38} &-23802.23
    \\
\midrule
\multicolumn{12}{c}{$\varepsilon_t \sim N \times t (\alpha,k)$}             \\
\hline
&$p$   & 1  & 2  & 3  & 4 & 5  &6 &7 &8 &9 &10 \\
\hline
\multirow{2}{*}{FTSE 100}  &MIC1  & 12716.15          & 12837.14      & 12828.04      & 12812.87          & 12821.58  & \underline{12866.07}          & 12787.31     & 12862.53    & 12746.68 & 12754.23 \\
&MIC2 & 12706.13 &\underline{12816.92}    & 12797.73     & 12772.53          & 12771.13  & 12805.35          & 12716.93      & 12781.66     & 12656.55          & 12654.07\\
\hline 
\multirow{2}{*}{Nikkei 225}  &MIC1 & 7589.51 &7593.72 &7602.29 &7590.82 &7595.97 &\underline{7606.32} &7545.75 &7568.10 &7587.28 &7602.21
  \\
&MIC2 &\underline{7579.88} &7574.47 &7573.40 &7552.38 &7547.92 &7548.61 &7479.00 &7491.63 &7501.09 &7506.32
    \\
\hline    
\multirow{2}{*}{S\&P 500}  &MIC1  & \underline{2404.78}        & 2348.69     & 2395.95     & 2132.21         &2189.68  &2339.97          & 2330.62     & 2305.78    &2198.37 & 2381.86 \\
&MIC2 &\underline{2400.86} &2341.04    & 2384.24     & 2118.33          & 2171.88  & 2317.16          & 2304.13      & 2275.85     & 2166.30          &2343.28\\
\bottomrule
\end{tabular}
}
\end{table}

\begin{table}[h]
\footnotesize
\caption{Parameter estimate results for the selected AR model.}\label{tr2}
\vspace{10pt}
\resizebox{\textwidth}{!}{
\begin{tabular}{lllllll}
\toprule
\multicolumn{6}{c}{$\varepsilon_t \sim N(0,\sigma^2)$}                                                      \\
\hline
&Order (method) & $(a_1,\cdots,a_p)$                                                                  & $c$   &$\sigma^2$\\
\hline
\multirow{2}{*}{FTSE 100} &$p=7$ (AIC)      &\makecell[l]{(-0.0004, -0.0266, -0.0448,  0.0426, -0.0215, -0.0406,  0.0308)} & 0.0002 &0.0001  \\
&$p=6$ (BIC)      &\makecell[l]{(-0.0016, -0.0272, -0.0435,  0.0412, -0.0223, -0.0406)}        &  0.0002   &0.0001   \\
\hline
\multirow{1}{*}{Nikkei 225} &$p=1$ (AIC, BIC)      &\makecell[l]{(-0.0309)} & 0.0003 &0.0002  \\
\hline
\multirow{1}{*}{S\&P 500} &$p=9$ (AIC, BIC)      &\makecell[l]{(-0.0926,  0.0487, -0.0177, -0.0449, -0.0143, -0.0601,  0.0795, -0.0580,  0.0594)} & 0.0004 &0.0001  \\
\midrule
\multicolumn{6}{c}{$\varepsilon_t \sim N \times t (\alpha,k)$}                                                      \\
\hline
&Order (method) & $(a_1,\cdots,a_p)$                                                                   & $c$ &s & $\alpha$ &$k$\\
\hline
\multirow{2}{*}{FTSE 100} &$p=6$ (MIC1)    &\makecell[l]{(-0.0212, -0.0813, -0.0158, -0.0265, -0.0230,
       -0.0331)} &0.0006 &0.0129 & 0.0104   & 2.0799  \\
&$p=2$ (MIC2)    & (-0.0256, -0.0877)                                                                        &0.0007 &0.0129 & 0.0104   & 2.0807\\
\hline
\multirow{2}{*}{Nikkei 225} &$p=6$ (MIC1)    &\makecell[l]{(-0.0445, -0.0020, -0.0260, -0.0157, -0.0303, 0.0031)} &0.0008 &0.0115 &  0.1030   & 1.2176 \\
&$p=1$ (MIC2)    & (-0.0479)                                                                        &0.0011 &0.0116 & 0.1052   & 1.2292\\
\hline
\multirow{1}{*}{S\&P 500} &$p=1$ (MIC1, MIC2)    &\makecell[l]{(-0.0168)} &0.0007 &0.0525 & 0.0093   & 1.9206  \\
\bottomrule
\end{tabular}
}
\end{table}

Since AR models offer a convenient framework for prediction, Table \ref{t-MSE} presents a comparison of out-of-sample performance between the fitted models, either chosen by AIC vs BIC or by MIC1  vs MIC2. 
In each case, we use the first part of the dataset for fitting and the remaining 100 data for the rolling $m$-step-ahead forecast. 
It is interesting to observe that, for the FTSE 100, the fitted AR(6) model selected by MIC1 performs better in forecasting for the short term than the fitted AR(2) model selected by MIC2 but worse for the longer term. However, for AIC vs BIC, the fitted AR(6) performs better than the fitted AR(7) uniformly. The results suggest that the effect of heavy-tailed innovation kicks in for the longer term forecasting. 
For the Nikkei 225, the fitted AR(6) model chosen by MIC1 achieves the best forecasting performance.
For the S\&P 500, the fitted AR(1) model selected by both MIC1 and MIC2, which involves heavy-tailed innovation, uniformly outperforms the fitted AR(9) model selected by AIC and BIC.

\begin{table}[htbp]
\footnotesize
\caption{MSE and the ratio in the rolling m-step-ahead forecast on AR models. (Row-wise minimum values are underlined.)}\label{t-MSE}
\vspace{10pt}
\resizebox{\textwidth}{!}{
\begin{tabular}{llcccccccc}
\toprule
Data &$m$  &   & \multicolumn{3}{c}{$\varepsilon_t \sim N \times t (\alpha,k)$} & & \multicolumn{3}{c}{$\varepsilon_t \sim N (0, \sigma^2)$} \\ 
\midrule
& & & $MSE (AR(2))$  & $MSE (AR(6))$ & $\frac{MSE (AR(2))}{MSE (AR(6))}$ &  & $MSE (AR(6))$  & $MSE (AR(7))$ & $\frac{MSE (AR(6))}{MSE (AR(7))}$\\
\cline{4-6} \cline{8-10} 
\multirow{5}{*}{FTSE 100} &1&                                      & $8.2026\times10^{-5}$  & $8.1961\times10^{-5}$  & 1.0008 &  & \underline{$8.1247\times10^{-5}$}  & $8.1534\times10^{-5}$  & 0.9965\\
& 2&                                     &$8.2508\times10^{-5}$  & $8.2340\times10^{-5}$ & 1.0020  & & \underline{$8.1492\times10^{-5}$}  & $8.1763\times10^{-5}$  & 0.9967\\
&    3&                                      & $8.1973\times10^{-5}$  & \underline{$8.1565\times10^{-5}$}  & 1.0050 &   & $8.1865\times10^{-5}$  & $8.2250\times10^{-5}$  & 0.9953\\
& 4&                                      & \underline{$8.1300\times10^{-5}$}  & $8.2374\times10^{-5}$  & 0.9870 &   & $8.3796\times10^{-5}$  & $8.4129\times10^{-5}$  & 0.9960\\
& 5&                                      & \underline{$8.2529\times10^{-5}$}  & $8.3208\times10^{-5}$  & 0.9918 &   & $8.4179\times10^{-5}$  & $8.4545\times10^{-5}$  & 0.9957\\
\midrule
& & & $MSE (AR(1))$  & $MSE (AR(6))$ & $\frac{MSE (AR(1))}{MSE (AR(6))}$ &  & $MSE (AR(1))$  &- &- \\
\cline{4-6} \cline{8-10} 
\multirow{5}{*}{Nikkei 225} &1&                                      & $4.3790\times10^{-4}$  & \underline{$4.3024\times10^{-4}$}  & 1.0178 &  & $4.3970\times10^{-4}$ &- &-\\
& 2&    &$4.4348\times10^{-4}$  & \underline{$4.4128\times10^{-4}$} & 1.0050  & &$4.4373\times10^{-4}$ &- &-\\
&    3&                                      & $4.4705\times10^{-4}$  & \underline{$4.4556\times10^{-4}$}  & 1.0033 &   &$4.4671\times10^{-4}$ &- &-\\
& 4&                                      & $4.5225\times10^{-4}$  & \underline{$4.4758\times10^{-4}$}  & 1.0104 &   &$4.5128\times10^{-4}$ &- &-\\
& 5&                                      & $4.5423\times10^{-4}$  & \underline{$4.5263\times10^{-4}$}  & 1.0035 &   &$4.5522\times10^{-4}$ &- &-\\
\midrule
& & &$MSE (AR(1))$ &- &- &  & $MSE (AR(9))$  &- &- \\
\cline{4-6} \cline{8-10} 
\multirow{5}{*}{S\&P 500} &1&                                      & \underline{$8.8073\times10^{-5}$} &- &- &  &$8.8697\times10^{-5}$  &- &-\\
& 2&                                     &\underline{$8.9475\times10^{-5}$} &- &-  & &$8.9767\times10^{-5}$  &- &-\\
&    3&                                      & \underline{$9.0349\times10^{-5}$} &- &- &   & $9.1379\times10^{-5}$  &- &-\\
& 4&                                      & \underline{$9.1280\times10^{-5}$} &- &- &   & $9.2134\times10^{-5}$  &- &-\\
& 5&                                      & \underline{$9.2046\times10^{-5}$} &- &- &   & $9.4757\times10^{-5}$  &- &-\\
\bottomrule
\end{tabular}
}
\end{table}

\subsection{Car Data} \label{5.2}

\begin{figure}[h]
\begin{center}
\includegraphics[width=0.59\textwidth]{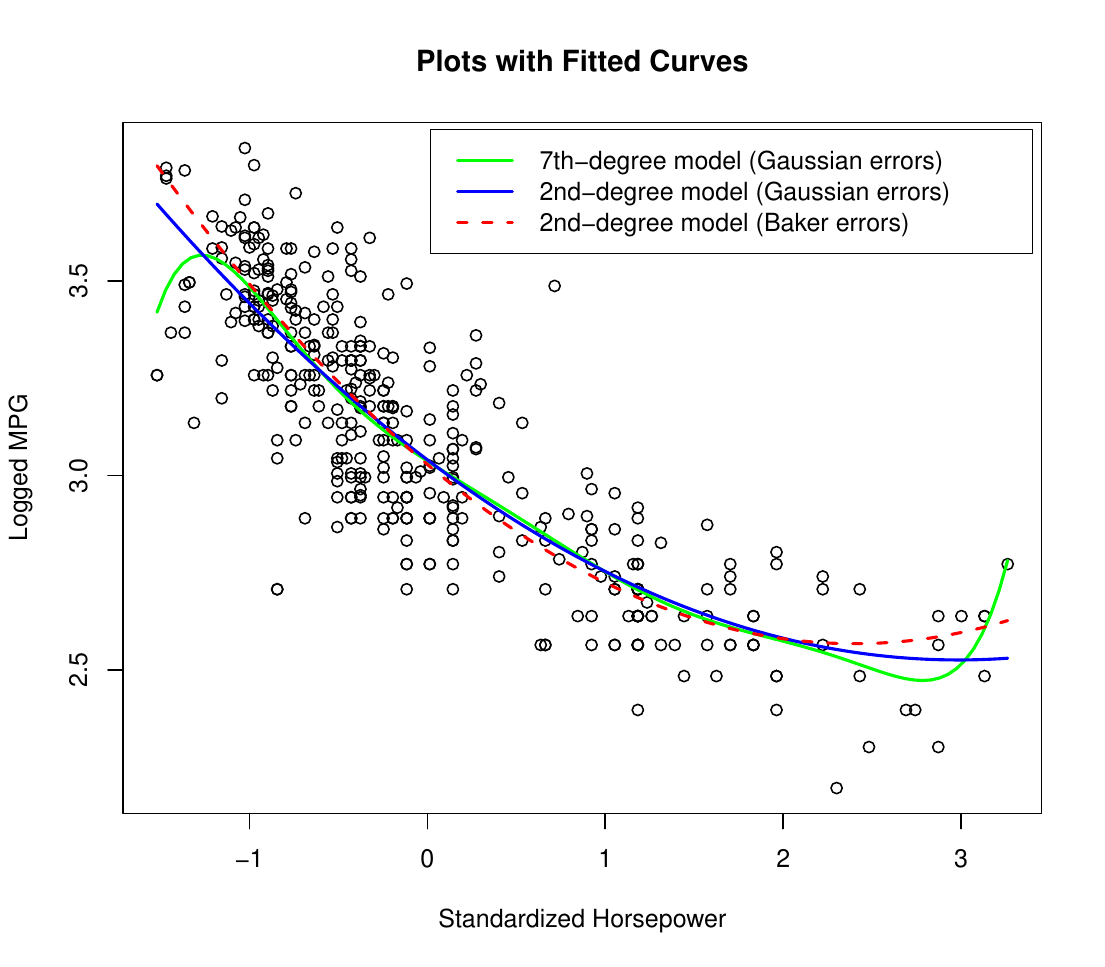}
\end{center}
\caption{Plots of logged mpg versus standardized horsepower and fits from different polynomial regression models.}\label{fig2}
\end{figure}

Here, we analyze the relationship between gas mileage in miles per gallon (mpg) and horsepower for 392 cars in the Auto dataset. This dataset, sourced from the StatLib library maintained at Carnegie Mellon University, was used in the 1983 American Statistical Association Exposition. 
Before the analysis, we apply a log transformation to mpg, as it is positive. 
To improve numerical stability, we also standardize horsepower. Figure \ref{fig2} shows the plot of logged mpg versus standardized horsepower and the various fitted polynomial regressions. 
 
The data plot suggests a  nonlinear relationship between mpg and horsepower. \cite{gareth2013introduction} fitted a polynomial regression model with Gaussian errors to this dataset. In this study, we apply a polynomial regression model with Baker errors and use the MIC1 and MIC2 to select the degree of the polynomial, setting candidate degrees from 1 to 10. For comparison, we also use AIC and BIC with Gaussian errors. For Baker errors we employ the model in equation (\ref{e_1_1}), and for Gaussian errors we employ the model given by
\begin{align}
    y=\beta_1x+\cdots+\beta_px^p+c+\varepsilon,\label{er-2}
\end{align}
where $\varepsilon \sim N(0,\sigma^2)$.

Tables \ref{tr3} and \ref{tr4} report model selection and parameter estimation, respectively. Under Gaussian errors, AIC selects a seventh-degree polynomial regression model, while BIC selects a quadratic polynomial regression model.
In contrast, both MIC1 and MIC2 select a quadratic polynomial regression model with Baker errors, and give estimates similar to those based on the BIC with Gaussian errors. This suggests that a quadratic  model chosen by BIC is an appropriate model rather than the seventh-degree model chosen by AIC. This example shows that MIC1 and MIC2 can be used to check the choice by AIC and BIC, even when the latter can be applied, and highlights an additional utility of MIC coupled with the Baker PDF.

\begin{table}[htbp]
\caption{Selection results for polynomial regression model with two error PDFs.}\label{tr3}
\vspace{10pt}
\resizebox{\textwidth}{!}{
\begin{tabular}{lllllllllll}
\toprule
\multicolumn{11}{c}{$\varepsilon \sim N(0,\sigma^2)$}\\
\hline
$p$   & 1  & 2  & 3  & 4 & 5 &6 &7 &8 &9 &10  \\
\hline
AIC & -186.29          & -242.96    & -243.44    & -242.01    & -249.08  & -248.42          & \underline{-250.44}    & -248.80    & -248.39    & -247.24  \\
BIC & -174.38          & \underline{-227.07}    & -223.59    & -218.18    & -221.29& -216.65          & -214.70    & -209.09    & -204.70    & -199.59    \\
\midrule
\multicolumn{11}{c}{$\varepsilon \sim N \times t (\alpha,k)$}             \\
\hline
$p$   & 1  & 2  & 3  & 4 & 5  &6 &7 &8 &9 &10 \\
\hline
MIC1 & 27.69          & \underline{34.24}    & 34.07    & 33.89    & 33.08 &29.71          & 32.66    & 28.84    & 32.38    & 31.93  \\
MIC2 & 27.41          & \underline{33.55}    & 33.05    & 32.55    & 31.45& 27.95          &30.42    & 26.60    & 29.56    & 28.86  \\
\bottomrule
\end{tabular}
}
\end{table}

\begin{table}[h]
\caption{Parameter estimate results for the selected polynomial regression model.}\label{tr4}
\vspace{10pt}
\resizebox{\textwidth}{!}{
\begin{tabular}{lllllll}
\toprule
\multicolumn{5}{c}{$\varepsilon \sim N(0,\sigma^2)$}                                                      \\
\hline
Degree (method) & $(\beta_1,\cdots,\beta_p)$                                                                 & $c$   &$\sigma^2$\\
\hline
$p=7$ (AIC)      &\makecell[l]{(-0.2966, 0.0723, -0.1232, 0.0412, 0.0489, -0.0321, 0.0051)} & 3.0376 &0.0296  \\
$p=2$ (BIC)      & (-0.3448,0.0578)                                                   & 3.0407   &0.0309   \\
\midrule
\multicolumn{5}{c}{$\varepsilon \sim N \times t (\alpha,k)$}                                                      \\
\hline
Degree (method) & $(\beta_1,\cdots,\beta_p)$                                                                 & $c$  &$s$ &$\alpha$ &$k$\\
\hline
$p=2$ (MIC1, MIC2)    &(-0.3838,  0.0800) & 3.0288 &0.3757 &0.4973 &3.2389\\
\bottomrule
\end{tabular}
}
\end{table}

We also examine the skewness and excess kurtosis of the residuals and consider the matching between the theoretical values and their sample estimates based on the fitted residuals, as shown in Table \ref{t-residual}.
Approximate standard errors (SE) of the sample estimates are obtained by 
a Bootstrap procedure, with 1000 replications. 
For the fitted models with Gaussian errors, the skewness and the excess kurtosis of the errors are both theoretically 0. On the other hand, for Baker errors, the skewness is theoretically 0, but the excess kurtosis would not be 0 unless $k=0$. Since it is difficult to calculate the excess kurtosis for Baker PDF due to its intractable normalizing constant,
we sample from Baker PDF and use the mean of the sample estimates from 1000 replications as the theoretical value, along with its SE.
Apparently, the fitted residuals of the models chosen by AIC, BIC, MIC1 and MIC2 have all produced very small negative skewness of similar size, matching their theoretical value of zero reasonably well. However, for excess kurtosis, while the matching is far from being satisfactory for the Gaussian models chosen by AIC and BIC, the excess kurtosis from the fitted model chosen by  MIC1 and MIC2 coupled with Baker PDF  is 1.29 (SE: 0.21) versus its residual counterpart of 1.33 (SE: 0.54), which is much better.

\begin{table}[h]
\caption{Comparison of residual skewness and excess kurtosis.}\label{t-residual}
\vspace{10pt}
\resizebox{\textwidth}{!}{
\begin{tabular}{lccccccc}
\toprule
Noise                     & Degree & & \multicolumn{2}{c}{Skewness} &  & \multicolumn{2}{c}{Excess kurtosis}     \\
\cline{4-5} \cline{7-8}
                          &    &   & Theoretical value& Sample estimate (SE) & & Theoretical value & Sample estimate (SE)\\
\midrule
\multirow{2}{*}{$N(0,\sigma^2)$} & $p=7$ & & 0           & -0.08 (0.22)          & & 0           & 1.13 (0.59)             \\
                          & $p=2$ & & 0           & -0.14 (0.20)          &  & 0           & 0.87 (0.49)             \\
\hline
$N \times t (\alpha,k)$                    & $p=2$ & & 0           & -0.22 (0.23)      &      & \textbf{1.29 (0.21)}        & \textbf{1.33 (0.54)}         \\
\bottomrule
\end{tabular}
}
\end{table}

\subsection{Wind Direction Data} \label{5.3}

\begin{figure}[htpb]
\begin{center}
\includegraphics[width=0.58\textwidth]{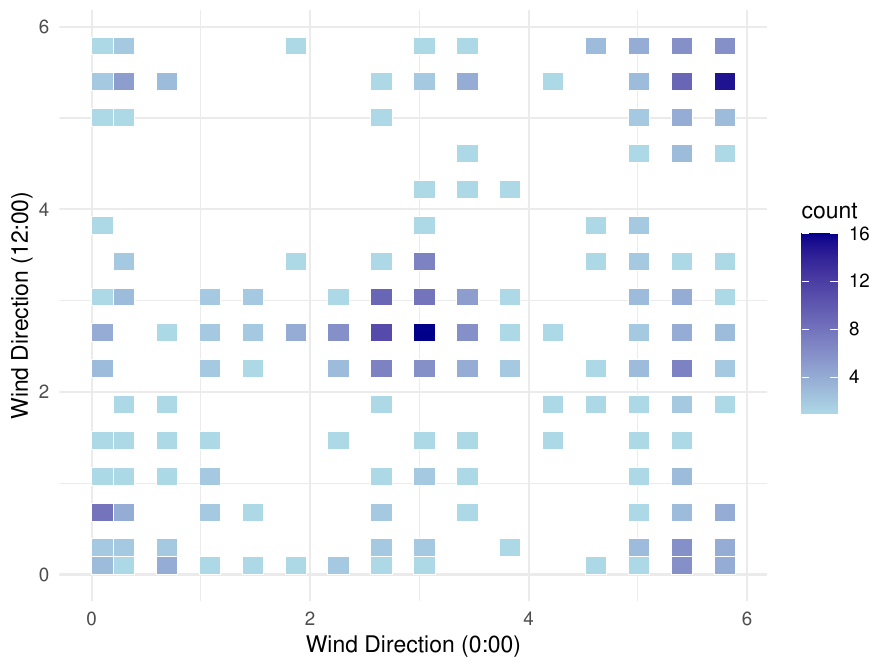}
\end{center}
\caption{2-D histogram of wind direction data.}\label{fig3}
\end{figure}

Finally, we fit a  bivariate model with von Mises PDF, as described in Section \ref{4.4}, to some wind direction data, using MIC1 and MIC2 for model selection. Here, the wind direction is represented as a circular variable in radians. \cite{matsuda2021information} applied this model to wind direction data from Tokyo at 00:00 and 12:00 in 2008. For the sake of cross-validation, we analyze more recent wind direction data from Tokyo at 00:00 ($x_1$) and 12:00 ($x_2$) over 365 days in 2023, obtained from the Japan Meteorological Agency website. The data are discretized into 16 bins, such as north-northeast. Figure \ref{fig3} presents the corresponding 2-d histogram. 

We fit the data using two candidate models, $m_1$ and $m_2$, representing two scenarios: $\lambda=0$ and $\lambda \in R$ in equation (\ref{e-von}), respectively. 
For comparison, results for MIC1, MIC2  are summarized in Table \ref{tr5}. For $m_2$, both MIC1 and MIC2 are higher,  indicating a better fit than $m_1$. This suggests that the wind directions at Tokyo on 00:00 and 12:00 in 2023 are dependent, consistent with the results in 2008 by \cite{matsuda2021information}.
\begin{table}[h]
\caption{Parameter estimate and MIC results for models $m_1$ and $m_2$.}\label{tr5}
\begin{center}
\begin{tabular}{lllllllll}
\toprule
Model &Order  & $\kappa_1$     & $\kappa_2$     & $\mu_1$    & $\mu_2$      & $\lambda$  &MIC1 &MIC2   \\
\midrule
$m_1$ &4             &0.4607 & 0.3057  & 5.6711 &2.3854   &0  &0.1014  &0.0971\\
$m_2$ &5             & 0.1872 & 0.1143   & 4.4243 &1.3127   &-1.5608  &\underline{0.9857} &\underline{0.9344}\\
\bottomrule
\end{tabular}
\end{center}
\end{table}

\section{Conclusion}\label{s6}
In this paper, we propose a consistent  selection criterion, MIC, for nested models with possibly unnormalized PDFs. Compared with existing model selection methods, MIC offers three advantages. First, it can handle both normalized and unnormalized PDFs. Second, it enjoys selection consistency for the case of a finite sequence of strictly nested models under mild regularity conditions. Third, it  reduces significantly computational costs by avoiding the calculation of the normalizing constant and the bias correction. 
To showcase the efficacy of MIC for unnormalized PDFs,
we have studied AR models and polynomial regression models with Baker noise/errors, the normalizing constants of which are typically intractable.  
Extensive simulation studies and real data applications have demonstrated consistency and effectiveness of MIC. Furthermore, we have shown excellent performance of MIC on PDFs with bounded support through experiments on models with bivariate von Mises PDFs and real wind direction data.

We have discussed how to use MIC for consistent model selection with PDFs supported on $\mathbb{R}^d.$ The simulation results and the real example with wind data strongly suggest that a promising direction for future research is to extend MIC to cover PDFs supported on bounded intervals  $(a,b)$. Another direction of extension is to discrete data, thereby availing MIC of the opportunity of selecting an appropriate model, such as an Ising model \citep{friel2013evidence,everitt2017bayesian},  in  the area of 
discrete Markov random fields and spatial statistics. It is also intriguing  to explore the use of combinations of different multiplying factors $C(n,k)$. 
We have set $C(n,k)$ to $\exp\{-2\#(M_k)/n\}$ and $n^{-\#(M_k)/n}$, respectively. Although $\exp\{-2\#(M_k)/n\}$ may not yield a consistent estimate of the true order, there is significant scope for combining it with $n^{-\#(M_k)/n}$, similar to existing approaches 
developed in \cite{ing2007}.
Last, but not least, an exploration of MIC into non-nested models should be exciting.

\section{Disclosure Statement}
We declare that we do not have any relevant financial or non-financial competing interests.

\bibliographystyle{agsm}

\appendix  
\section{Appendix}{\label{sapp}}
\subsection{Proof of Proposition \ref{prop1}}
\begin{proof}
To simplify notations, we define an average operator, $P_n$, for data sample $x_1,\cdots ,x_n$ applied to any function $g(x,\cdot)$ by
\[
P_n[g(x)]=\frac{1}{n}\sum_{i=1}^ng(x_i,\cdot).
\]
Therefore, we have that
\[
GIC(M(\theta))=P_n[W(x,p_{M(\theta)})].
\]
Let the constrained MGICE of $\alpha$ be $\hat{\alpha}_0$ while the unconstrained MGICE of $\alpha$ and $\beta$ be $\hat{\alpha}$ and $\hat{\beta}$, respectively. Note that $\hat{\alpha}_0$ satisfies the equation:
\begin{align}
    0 &= P_n\nabla_\alpha W(x, p_{M( \hat{\alpha}_0,\beta^*)}) \nonumber\\ 
      &= P_n\nabla_\alpha W(x, p_{M(\alpha^*,\beta^*)}) + P_n\nabla_\alpha \nabla^T_\alpha W(x, p_{M(\alpha^*,\beta^*)})(\hat{\alpha}_0 - \alpha^*) + O_p(n^{-1})
\end{align}
where the $O_p(n^{-1})$ term follows from mean value theorem and Assumption~\ref{ass8}, and that the MGICE is root-n consistent. Similarly, the unconstrained MGICE satisfies the following equation.
\begin{align}
0 = P_n\nabla_\alpha W(x, p_{M(\alpha^*,\beta^*)}) + P_n\nabla_\alpha\nabla^T_\alpha W(x, p_{M(\alpha^*,\beta^*)})(\hat{\alpha} - \alpha^*) \nonumber\\ 
+P_n\nabla_\alpha\nabla^T_\beta W(x, p_{M(\alpha^*,\beta^*)})(\hat{\beta} - \beta^*) + O_p(n^{-1}).
\end{align}
The preceding two equations imply that
\begin{align}
\hat{\alpha}_0 - \alpha^* = \hat{\alpha} - \alpha^* + D^{-1}_{\alpha^*,\alpha^*}D_{\alpha^*,\beta^*}(\hat{\beta} - \beta^*) +O_p(n^{-1}),
\end{align}
where $D_{a^*,b^*} = -P_n\nabla_a \nabla^T_b W(x, p_{M(\alpha^*,\beta^*)})$, with $a$, $b$ being either $\alpha$ or $\beta$, and $D_{\alpha^*,\alpha^*}$ is invertible. Doing a Taylor expansion around the constrained MGICE and after some algebra, we have
\begin{align}
&[\log(GIC(M(\alpha^*, \beta^*))) - \log P_n\{W(x, p_{M( \hat{\alpha}_0,\beta^*)})\}] \times [2 P_n\{W(x, p_{M(\alpha^*,\beta^*)})\}]\nonumber\\ 
&= -(\hat{\alpha}_0 - \alpha^*)^T D_{\alpha^*,\alpha^*} (\hat{\alpha}_0 - \alpha^*) + O_p(n^{-3/2}) \nonumber\\ 
&=-
\begin{pmatrix}
 (\hat{\alpha} - \alpha^*)^T & (\hat{\beta} - \beta^*)^T 
\end{pmatrix} 
\begin{pmatrix}
 I & 0\\
 D_{\beta^*,\alpha^*}D^{-1}_{\alpha^*,\alpha^*} &0
\end{pmatrix}
\begin{pmatrix}
 D_{\alpha^*,\alpha^*} & D_{\alpha^*,\beta^*}\\
 D_{\beta^*,\alpha^*}  & D_{\beta^*,\beta^*} 
\end{pmatrix}
\begin{pmatrix}
 I & D^{-1}_{\alpha^*,\alpha^*}D_{\alpha^*,\beta^*}\\
 0  & 0 
\end{pmatrix}
\binom{\hat{\alpha} - \alpha^*}{\hat{\beta} - \beta^*} 
\nonumber\\
&+O_p(n^{-3/2}) \nonumber\\
&=-
\begin{pmatrix}
 (\hat{\alpha} - \alpha^*)^T & (\hat{\beta} - \beta^*)^T 
\end{pmatrix}
\begin{pmatrix}
 D_{\alpha^*,\alpha^*} & D_{\alpha^*,\beta^*}\\
 D_{\beta^*,\alpha^*}  & D_{\beta^*,\alpha^*}D^{-1}_{\alpha^*,\alpha^*}D_{\alpha^*,\beta^*} 
\end{pmatrix}
\binom{\hat{\alpha} - \alpha^*}{\hat{\beta} - \beta^*} + O_p(n^{-3/2}) \label{AE1}
\end{align}
where $I$ denotes the identity matrix of order $h_{k_0}$. Similarly, we have
\begin{align}
&[\log(GIC(M(\alpha^*, \beta^*))) - \log P_n\{W(x, p_{M( \hat{\alpha},\hat{\beta})})\}] \times [2 P_n\{W(x, p_{M(\alpha^*,\beta^*)})\}]\nonumber\\ 
&=-
\begin{pmatrix}
 (\hat{\alpha} - \alpha^*)^T & (\hat{\beta} - \beta^*)^T 
\end{pmatrix}
\begin{pmatrix}
 D_{\alpha^*,\alpha^*} & D_{\alpha^*,\beta^*}\\
 D_{\beta^*,\alpha^*}  & D_{\beta^*,\beta^*} 
\end{pmatrix}
\binom{\hat{\alpha} - \alpha^*}{\hat{\beta} - \beta^*} + O_p(n^{-3/2}) \label{AE2}
\end{align}
Note that the matrix in the middle of the quadratic form depends on $n$ and it converges in probability to $D(\theta^*)$ defined in (\ref{matrixD}) which is a positive definite matrix. We can similarly partition it into a 2 by 2 block matrix:
\begin{align}
D(\theta^*)=
\begin{pmatrix}
 D(\alpha^*,\alpha^*) & D(\alpha^*,\beta^*)\\
 D(\beta^*,\alpha^*)  & D(\beta^*,\beta^*)
\end{pmatrix}  
\end{align}
Subtracting (\ref{AE2}) from (\ref{AE1}) yields:
\begin{align}
    &[\log P_n\{W(x, p_{M( \hat{\alpha},\hat{\beta})})\} - \log P_n\{W(x, p_{M( \hat{\alpha}_0,\beta^*)})\}] \times [2 P_n\{W(x, p_{M(\alpha^*,\beta^*)})\}]\nonumber\\ 
    &=(\hat{\beta} - \beta^*)^T(D_{\beta^*,\beta^*}-D_{\beta^*,\alpha^*}D^{-1}_{\alpha^*,\alpha^*}D_{\alpha^*,\beta^*})(\hat{\beta} - \beta^*) + O_p(n^{-3/2}) 
\end{align}
However, $\log P_n\{W(x, p_{M( \hat{\alpha},\hat{\beta})})\} - \log P_n\{W(x, p_{M( \hat{\alpha}_0, \beta^*)})\} = \log GIC(M_k(\hat{\theta}_k)) - \log GIC(M_{k_0}(\hat{\theta}_{k_0})).$ Recall $\sqrt{n}(\hat\theta - \theta^*)$ is asymptotically normally distributed with mean zero and covariance matrix equal to $D^{-1}(\theta^*)\Lambda (\theta^*)D^{-T}(\theta^*)$. It follows from routine algebra that $\sqrt{n}(\hat{\beta} - \beta^*)$ is asymptotically normal with zero mean vector and covariance matrix equal to $B(\theta^*)B^T(\theta^*)$.
Hence, $n\times [\log GIC(M_k(\hat{\theta}_k)) - \log GIC(M_{k_0}(\hat{\theta}_{k_0}))]$ converges in distribution to the non-negative random variable $Z^TA(\theta^*)Z$.
\end{proof}

\subsection{Proof of Theorem \ref{thm-TFIC-consistency}}
\begin{proof}
First, we observe that for any model $M$ satisfying Assumptions~\ref{ass6}--\ref{ass7}, it follows from the law of large numbers and routine analysis that $GIC(M(\theta)) = P_n\{W(x, p_{M(\theta)})\}$ converges uniformly in probability to its population version $GIC_\infty(M(\theta)) = E_{p^*}[W(x, p_{M(\theta)})]$, where $p^*$ is the true population PDF. It follow from Assumption~\ref{ass7} that $GIC_\infty(M_k(\theta_k))$ is a Lipschitz-continuous, as a function of the parameter $\theta_k$, hence it attains its maximum value, denoted by $\mathcal{M}_k$, owing to the compact parameter space assumption (aka Assumption~\ref{ass6}). Since $k_0$ is the smallest $k$ such that $M_k$ contains the true model, it follows that $ \mathcal{M}_k < \mathcal{M}_{k_0}$ for all $1 \le k < k_0$, whereas $\mathcal{M}_k = \mathcal{M}_{k_0}$ otherwise. Note that from Proposition 3 in \cite{cheng2024foundation}, $\mathcal{M}_{k_0} = H_G (p^*) > 0 $. Therefore, if for any $k$, $C(n, k) \to  1$ as $n \to  \infty$, then the maximum MIC model selection criterion will not select any $k < k_0$, in probability. 

Henceforth, consider the case that $k \ge k_0$. Since $\mathcal{M}_{k_0} > 0$, $GIC(M_k(\hat{\theta}_k))$ is positive, in probability, i.e., $GIC(M_k(\hat{\theta}_k)) > 0$ holds with probability approaching 1 as sample size increases without bound. For ease of exposition, we shall assume that $GIC(M_k(\hat{\theta}_k))$ is positive. Let $k > k_0$ be fixed. Consider the increment $D = \log\{MIC(k)\} - \log\{MIC(k_0)\} = \log C(n, k) - \log C(n, k_0) + \log GIC(M_k(\hat{\theta}_k)) - \log GIC(M_{k_0}(\hat{\theta}_{k_0}))$.
By Proposition \ref{prop1}, $n\times \{\log GIC(M_k(\hat{\theta}_k)) - \log GIC(M_{k_0}(\hat{\theta}_{k_0}))\}$ converges weakly to some non-negative random variable. 
Consequently, $ D = \log C(n, k) - \log C(n, k_0) + O_p(1/n)$ so that if $n \times \log\{C(n, k)/C(n, k_0)\} \to  -\infty$ as $n \to  \infty$, $D$ is negative in probability for $k > k_0$. 
This completes the proof of the consistency of the proposed MIC model selection criterion.
\end{proof}

\end{document}